\documentclass[a4paper,11pt]{article}

\usepackage[utf8]{inputenc}
\usepackage[numbers,sort&compress]{natbib}

\usepackage[margin=1.0in]{geometry}

\usepackage{latexsym}
\usepackage{graphicx}
\usepackage{color}
\usepackage{amsmath,amssymb}
\usepackage{hyperref}
\usepackage{tikz}
\usepackage{subcaption}
\usepackage{float}
\usepackage{soul}
\allowdisplaybreaks

\newcommand{\re}{\operatorname{Re}}
\newcommand{\im}{\operatorname{Im}}
\newcommand{\be}{\begin{equation}}
\newcommand{\ee}{\end{equation}}
\newcommand{\bpm}{\begin{pmatrix}}
\newcommand{\epm}{\end{pmatrix}}
\newcommand{\mrm}{\mathrm}

\begin{document}

\thispagestyle{empty}

 \hfill CP3-Origins-2019-33 DNRF90

\vfill
\begin{center}
  {\Large {\boldmath\bf {Higgs-Boson Masses and Mixings in the MSSM with CP~Violation and Heavy SUSY Particles}
\par} \vskip 2.5em
{\large
{\sc Nick Murphy$^{1}$, Heidi Rzehak$^{1,2}$
     }\\[2ex]
{\normalsize \it 
$^1$ $\text{CP}^3$-Origins, University of Southern Denmark, \\
Campusvej 55, DK-5230 Odense M, Denmark\\[3mm]
$^2$ Institute for Theoretical Physics, University of T\"ubingen,\\{ Auf der Morgenstelle 14, 72076 T\"ubingen, Germany}}
}}
\par \vskip 1em
\end{center}\par
\vskip .0cm \vfill {\bf Abstract:} We calculate the Higgs-boson mass spectrum
and the corresponding mixing of the Higgs states in the Minimal Supersymmetric
Standard Model (MSSM). We assume a mass-hierarchy with heavy SUSY particles and
light Higgs bosons. To investigate this scenario, we employ an
effective-field-theory approach with a low-energy Two-Higgs-Doublet Model
(2HDM) where both Higgs doublets couple to up- as well as down-type fermions.
We perform a one-loop matching of the MSSM to the 2HDM and evolve the
parameters to the low energy scale by exploiting two-loop renormalization group
equations, taking the complex parameters into account. For the calculation of
the pole mass, we compare three different options: one suitable for large
charged Higgs masses, one for low charged Higgs masses, and one approximation
that interpolates between these scenarios. The phase dependence of the mass
of the lightest neutral Higgs boson can be sizeable, i.e.\ on the order of a
couple of GeV depending on the scenario. In addition, we discuss the CP
composition of the neutral Higgs bosons. 
\setcounter{page}{0}
\clearpage
\def\thefootnote{\arabic{footnote}}
\setcounter{footnote}{0}

\section{Introduction}
\label{se:intro}
\paragraph{}
Supersymmetric extensions of the Standard Model (SM) can help overcome shortcomings of the SM, providing for example dark matter
candidates and additional sources of CP violation. While these models are theoretically attractive, none of the
predicted supersymmetric (SUSY) particles has been found thus far. In order to guide experimentalists in their
continued hunt for these yet-undiscovered states and to ascertain the continued theoretical viability of these theories,
precise theoretical predictions of experimentally observable quantities are needed. 

The discovery of the Higgs boson in 2012 \cite{Aad:2012tfa, Chatrchyan:2012xdj} has provided theorists with a whole new
set of results against which these theories may be tested. The best measured property of the Higgs boson is its mass
$m_h$ with a value of 
\begin{align}
m_h = 125.09 \pm 0.24 \text{ GeV} \quad \text{\cite{Aad:2015zhl}}.
\end{align} 
This mass is a free parameter in the SM, and must be determined from experiment. In supersymmetric extensions,
however, the mass of the discovered Higgs boson can be predicted using the additional parameters of the theory.
Demanding that the theoretical prediction match the experimentally-measured value for each point in the parameter space
constrains the theory. Since the experimental result for the Higgs mass is very precise, it is necessary to obtain a
theoretical prediction that is of similar accuracy in order to fully exploit the information of the experiment and
to yield stringent constraints. This entails incorporating quantum corrections in the Higgs mass calculations. 

The necessity of including quantum corrections of higher order in the Higgs-mass prediction has been recognized for a
long time. In the context of the Minimal Supersymmetric Standard Model (MSSM), radiative corrections have been shown to be important to lift the mass of the lightest Higgs boson above the
tree-level upper limit, which is given by the Z-boson mass \cite{Haber:1990aw, Ellis:1990nz, Okada:1990vk,
Ellis:1991zd}. A lot of work has been performed to improve the theoretical prediction: At fixed order, one-loop
\cite{Chankowski:1991md, Brignole:1992uf, Chankowski:1992er, Dabelstein:1994hb, Pierce:1996zz, Frank:2006yh}, two-loop
\cite{Heinemeyer:1998jw, Heinemeyer:1998kz, Zhang:1998bm, Heinemeyer:1998np, Espinosa:1999zm,  Espinosa:2000df,  Degrassi:2001yf,
Brignole:2001jy, Brignole:2002bz, Martin:2002iu, Martin:2002wn, Dedes:2002dy, Dedes:2003km, Martin:2004kr, Allanach:2004rh,
Heinemeyer:2004xw, Martin:2005eg, Heinemeyer:2007aq, Borowka:2014wla, Degrassi:2014pfa, Hollik:2014wea, Hollik:2014bua,
Hollik:2015ema, Borowka:2015ura, Goodsell:2016udb, Passehr:2017ufr, Borowka:2018anu} as well as three-loop corrections
\cite{Martin:2007pg, Harlander:2008ju, Kant:2010tf, Harlander:2017kuc, Stockinger:2018oxe, R.:2019ply, R.:2019irs} to the Higgs-boson masses have been
calculated. As it stands, the theoretical uncertainty of the lightest Higgs boson has been estimated to be of the order 3 GeV
~\cite{Degrassi:2002fi}, an estimate which is still applied in phenomenological studies. While the assessment of the
theoretical uncertainty is still an ongoing discussion (see Refs.~\cite{Vega:2015fna, Bahl:2017aev, Allanach:2018fif, Bahl:2019hmm, Slavich:2020zjv,Domingo:2021kud, R:2021bml}),
the general consensus is that it is challenging to reduce the uncertainty below 1 GeV.

The method by which one includes these quantum corrections is dependent on  the mass of  the SUSY particles. The
aforementioned fixed-order approach is particularly useful for SUSY particles with masses up to the TeV scale where the
mass of the SUSY partners of the top quark is most important. SUSY particles at this scale have been motivated by naturalness and grand-unification arguments among others, but as there are still no signs of low-energy SUSY at the LHC, 
heavy SUSY particles have begun to attract more interest. Within these scenarios, the fixed-order calculations provide a
less accurate result due to the emergence of large logarithms of ratios of masses of SUSY particles and the energy scale
at which the calculation is performed.  These large logarithms spoil the fixed-order perturbation series. In order to
take these large logarithms into account, an effective-field theory approach has been applied.
In this approach, the full MSSM governs the interaction behaviour at the
high-energy scale, and the effects of the heavy SUSY particles on the
low-energy physics are encoded into the couplings of a viable low-energy theory
such as the SM or the Two-Higgs-Doublet Model (2HDM) via matching at the
matching scale. This matching entails calculating the couplings in both the
high and low energy theories up to fixed order such that the physics described
by the MSSM and the 2HDM at the matching scale is the same. The resulting
couplings are then evolved down to the low-energy scale at which
the observables are investigated with the help of the corresponding renormalization group equations (RGE). Solving the
renormalization group equations leads to a resummation of the large logarithms.  Assuming the SM as low-energy theory,
leading logarithms (LL) have been resummed in Ref.~\cite{Barbieri:1990ja, Okada:1990gg} and next-to leading logarithms
(NLL) in Ref.~\cite{Kodaira:1993yt, Hempfling:1993qq, Casas:1994us, Haber:1996fp} taking into account top-Yukawa
corrections.  Allowing all Higgs bosons to be light, LL corrections  have been calculated in Ref.~\cite{Espinosa:1991fc,
Sasaki:1991qu, Chankowski:1992ek, Haber:1993an}.  Analytical expressions for the Higgs-boson mass have been obtained
taking LL contributions into account up to two-loop order~\cite{Carena:1995bx}. A hierachical stop-mass spectrum with
one stop-quark mass much heavier than the other has been considered in Ref.~\cite{Espinosa:2001mm} and the resulting
two-loop LL and NLL corrections of the Higgs-boson mass have been calculated. This approach has also been used to obtain
one- and two-loop LL corrections to the Higgs-boson mass spectrum in a CP-violating scenario \cite{Pilaftsis:1999qt, Carena:2000yi}. In
Ref.  \cite{Draper:2013oza}, this approach has been performed taking into account an intermediate step with light
electroweak fermionic SUSY partners and the SM as low-energy theory. The results have been improved further allowing
also for light Higgs bosons in Ref.~\cite{Lee:2015uza, Bagnaschi:2015pwa}\footnote{Matching conditions of the MSSM to the 2HDM are also discussed in Ref.~\cite{Gorbahn:2009pp}.}. The calculations of Refs.~\cite{Draper:2013oza, Lee:2015uza} have been implemented into the tool
MhEFT. A further tool was presented with SUSYHD~\cite{Vega:2015fna}, which includes threshold corrections to a higher
order. The threshold correction to the quartic Higgs coupling, assuming the SM as low-energy theory, has been calculated
taking into account two-loop QCD~\cite{Bagnaschi:2014rsa}, top Yukawa~\cite{Bagnaschi:2017xid}, both in the gaugeless limit, and  the full QCD corrections~\cite{Bagnaschi:2019esc}. Effects
of higher-dimensional operators have been studied in Refs.~\cite{Bagnaschi:2017xid, Wells:2017vla} where
Ref.~\cite{Wells:2017vla} exploits a different method to obtain the one-loop threshold corrections.  The
prediction of the Higgs-boson mass has been improved by resumming logarithms of fourth logarithmic order
($\text{N}^3$LL)~\cite{Harlander:2018yhj} in the  case that the SM is the low-energy theory.  Furthermore, a combination
of both approaches, the fixed-order and the RGE approach, has been performed, in particular to improve the intermediate
regime with particles not very heavy but too heavy for the existing fixed-order results~\cite{Hahn:2013ria,
Bahl:2016brp,  Bahl:2018jom, Bahl:2019wzx, Bahl:2020tuq}. These results are implemented in FeynHiggs~\cite{Frank:2006yh, Heinemeyer:1998np,
Degrassi:2002fi, Hahn:2013ria, Bahl:2016brp, Heinemeyer:1998yj, Bahl:2018qog}. Similarly,
FlexibleSUSY~\cite{Athron:2014yba, Athron:2017fvs} as well as a version of Sarah/SPheno~\cite{Staub:2009bi,
Staub:2010jh, Staub:2012pb,  Staub:2013tta, Porod:2003um, Porod:2011nf} have implemented such a
combination~\cite{Athron:2016fuq, Staub:2017jnp}.  Furthermore, in Ref.~\cite{Harlander:2019dge},  $\text{N}^3$L0-fixed-order and $\text{N}^3$LL results have been combined to give a precise prediction for the mass of the lightest Higgs boson. A review about the different Higgs-mass calculations can be found in Ref.~\cite{Slavich:2020zjv}.

The particular theory we wish to explore in this work is the  
MSSM with complex parameters where the SUSY particles  are heavy and the additional Higgs bosons have masses at an intermediate scale. In addition to studying the mass of
the lightest neutral Higgs boson in this scenario, we analyze the size of this boson's CP-odd component, which is
induced by quantum corrections in the presence of complex parameters of the high-energy theory. We  also look into the
mixing and masses of the heavy Higgs bosons. For this, we assume a type-III complex 2HDM as the low-energy effective
theory, where both Higgs doublets can couple to up- as well as down-type fermions and use different approaches  to
connect to the Standard Model.  For vanishing bottom-Yukawa couplings, these results are implemented in FeynHiggs as the EFT part of the hybrid result discussed in Ref.~\cite{Bahl:2020mjy}, additionally improved by gauge-coupling and two-loop contributions in the matching conditions using results from Refs.~\cite{Bahl:2020tuq, Lee:2015uza} (see also Ref.~\cite{Bahl:2020jaq} for two-loop corrections).

Other studies of a CP-violating MSSM have been done. The  scan of the phenomenological MSSM in Ref.~\cite{Arbey:2014msa}
did not find a measurable size of the CP-odd component of the lightest Higgs boson, while the study presented in
Ref.~\cite{Li:2015yla} found some more promising results with some scenarios that could be measurable at least at future
runs of the LHC. In Ref.~\cite{Carena:2015uoe} a scenario with heavy superpartners has been explored
using complex MSSM parameters to include CP-violating effects, taking into account the finding of Ref.~\cite{Lee:2015uza}.

In this paper, we further improve the results for scenarios of an MSSM with complex parameters and heavy superpartners: We exploit two-loop RGEs 
allowing for complex parameters  and a one-loop matching of the MSSM to the 2HDM type III leading to a result where the NLL are resummed.  We will start out with setting our conventions for the MSSM in Section \ref{se:MSSM}
and continue to describe the considered low-energy theory in Section \ref{se:efftheory}. In Section \ref{se:matching} we
discuss the exploited matching procedure and in Section \ref{se:mass}, we present the details of the determination of
the Higgs-boson masses as well as their mixing. The numerical results are discussed in Section~\ref{se:results}, and,
finally, we conclude in Section \ref{se:conclusion}.


\section{The Minimal Supersymmetric Standard Model} \label{se:MSSM}
\paragraph{}
As mentioned in Section \ref{se:intro}, we consider a scenario in which all the superpartner particles are much heavier
than the SM particles and the Higgs bosons. Hence, the full MSSM is active at a high-energy scale and the effects of the
superpartners enter into the low-energy theory via threshold effects. The aim of this section is to set up the notation
needed for the calculation of the threshold corrections.

The superpotential of the MSSM is given as
\begin{align}\label{superpotential}
	W_{\text{MSSM}} = - \epsilon_{ij}\left[h_u^{\text{MSSM}} \check{H}_u^i \check{Q}^j \check{U}^c - h_d^{\text{MSSM}}
	\check{H}_d^i \check{Q}^j \check{D}^c -h_{e}^{\text{MSSM}}\check{H}_d^i \check{L}^j \check{E}^c + \mu
	\check{H}_d^i\check{H}_u^j\right],
\end{align}
where all fields denoted with a $\check{}$ are left-chiral superfields. $\check{Q}$ and $\check{L}$ denote the quark and lepton
superfield doublets, $\check{U}$, $\check{D}$ and $\check{E}$ denote the up-type quark, down-type quark, and lepton
charge-conjugate superfield singlets, the two Higgs-doublet superfields are denoted by $\check{H}_u$ and $ \check{H}_d$, and $\epsilon_{12}
= 1$. The
corresponding Yukawa couplings are $h_u^{\text{MSSM}}$, $h_d^{\text{MSSM}}$, and $h_{e}^{\text{MSSM}}$, which are
complex $3\times3$ matrices in general. However, in our calculation, we will neglect the first two generations, and the
matrices collapse to the the top-, bottom-, and tau-Yukawa coupling $h_t^{\text{MSSM}}$, $h_b^{\text{MSSM}}$, and
$h_{\tau}^{\text{MSSM}}$, which can be chosen to be real \cite{Kobayashi:1973fv}. The strength of the mixing of the two Higgs doublets
is described by the complex parameter $\mu$.

We do not explicity give the vector part of the Lagrangian with the kinetic and interaction terms for the gauge bosons,
but refer the reader to e.g. Refs.~\cite{Martin:1997ns, Drees:1996ca}.

Since supersymmetry cannot be exact, it is explicitly broken in the MSSM by soft-SUSY breaking terms:
\begin{align}\label{soft_breaking}\nonumber
	\mathcal L_{\text{MSSM}}^{\text{soft}} &= - m_{H_d}^2 |H_d|^2 - m_{H_u}^2 |H_u|^2 
	\\\nonumber
	&- M_{L_Q}^2 |\tilde{Q}|^2  - M_{R_U}^2 |\tilde{u}_R|^2  - M_{R_D}^2 |\tilde{d}_R|^2 
- M_{L_L}^2 |\tilde{L}|^2 - M_{R_E}^2 |\tilde{e}_R|^2 \\& \quad \nonumber
+ \epsilon_{ij} (m_{H_dH_u}^2 H_d^i H_u^j + h_u^{\text{MSSM}} A_u H_u^i \tilde{Q}^j \tilde{u}^*_R 
                -  h_d^{\text{MSSM}} A_d H_d^i \tilde{Q}^j \tilde{d}^*_R \\& \qquad \qquad \nonumber
                -  h_e^{\text{MSSM}} A_e H_d^i \tilde{L}^j \tilde{e}^*_R + \mathrm{h.c.})\\& \quad
- \frac{1}{2} (M_1  \tilde{B} \tilde{B} +   M_2 \tilde{W}_i \tilde{W}_i + M_3
\tilde{G} \tilde{G} + \mathrm{h.c.}) \; .
\end{align}
where $\tilde{Q}$, $\tilde{L}$, $\tilde{U}$, $\tilde{D}$, $\tilde{E}$, $H_d$ and $H_u$ denote the scalar component of
the corresponding superfields. The $\tilde{}$ indicates a superpartner field.  The gaugino fields corresponding to
$U(1)$, $SU(2)$, and $SU(3)$  are denoted by $\tilde{B}$, $\tilde{W}$, and $\tilde{G}$, respectively. We assume colour
indices to be implicit. The gaugino soft breaking parameters $M_1$, $M_2$, and $M_3$ as well as the Higgs mixing
parameter $m_{H_dH_u}^2$ are complex numbers, while the soft Higgs mass breaking parameters $m_{H_d}^2$, $m_{H_u}^2$
are real. In general, the sfermion mass parameters $M_{L_Q}^2$, $M_{R_U}^2$, $M_{R_D}^2$, $M_{L_L}^2$, $M_{R_E}^2$ are
$3\times3$ Hermitian matrices, but reduce to real parameters when generation mixing is ignored. Finally, the trilinear
couplings $A_u$, $A_d$, and $A_e$ are general $3 \times 3$ complex matrices, but reduce to complex numbers if they are
assumed to be proportional to the SM Yukawa matrices, as is done in this paper.

In this paper, we are concerned with the Higgs sector of the MSSM. Equations \ref{superpotential} and
\ref{soft_breaking} give rise to a Higgs potential of the form 
\begin{multline}\label{higgs_potential}
V_H = \frac{1}{8}(g^2+g_y^2)(|H_d|^2-|H_u^2|)^2 + \frac{1}{2}g^2|H_d^\dagger H_u|^2+|\mu|^2(|H_d|^2+|H_u|^2) \\
	+ m_{H_d}^2|H_d|^2 + m_{H_d}^2|H_u^2| - m_{H_dH_u}^2(\epsilon_{ab}H^a_d H^b_u + \mathrm{h.c.}) 
\end{multline}
for the two Higgs doublets $H_u$ and $H_d$ of hypercharge 1 and -1, respectively. The  $SU(2)$ and the $U(1)$ gauge coupling are denoted by $g$ and $g_y$, respectively. 
Finally, the squark mass matrices are given by 
 \begin{align}\label{Sfermionmassenmatrix}
\mathcal M_{\tilde{q}} &= \begin{pmatrix} 
M_{L_Q}^2 + m_q^2 + M_Z^2 c_{2 \beta} (T_q^3 - Q_q \sin^2 \theta_W) & 
 m_q X_q^* \\[.2em]
 m_q X_q &
 M_{R_F}^2 + m_q^2 +M_Z^2 c_{2 \beta} Q_q \sin^2 \theta_W
\end{pmatrix}
\end{align}
with
\begin{align}\label{kappa}
X_q &= A_q - \mu^*\kappa~, \quad \kappa = \{\cot\beta, \tan\beta\}  \text{ and } F = \{U, D\}
        \quad {\rm for} \quad q = \{t, b\}~.
\end{align}
Here, we introduce the gauge-boson mass $M_Z$, the electroweak mixing angle $\theta_W$, the quark masses $m_q$, as well
as $\beta$, which is defined via the ratio of the Higgs vacuum expectation values of the MSSM, $\tan \beta \equiv
v_u/v_d$. The charge and the third component of the isospin of the squarks are denoted by $Q_q$ and $T_q^3$,
respectively.


\section{The Effective Low-energy Theory}
\label{se:efftheory}
\paragraph{}
The resulting low-energy theory is a Two-Higgs-Doublet Model (2HDM) with the following Higgs
potential $V$
\begin{align}\label{THDM_potential}
V &= m_{11}^2\Phi_1^{\dagger}\Phi_1 + m_{22}^2\Phi_2^{\dagger}\Phi_2 - [m_{12}^2\Phi_1^{\dagger}\Phi_2 +
	\textrm{h.c.}]\\\nonumber & + \frac{1}{2}\lambda_1(\Phi_1^{\dagger}\Phi_1)^2 +
	\frac{1}{2}\lambda_2(\Phi_2^{\dagger}\Phi_2)^2 + \lambda_3(\Phi_1^{\dagger}\Phi_1)(\Phi_2^{\dagger}\Phi_2) +
	\lambda_4(\Phi_1^{\dagger}\Phi_2)(\Phi_2^{\dagger}\Phi_1)\\\nonumber &  +
	\big\{\frac{1}{2}\lambda_5(\Phi_1^{\dagger}\Phi_2)^2  +  [\lambda_6(\Phi_1^{\dagger}\Phi_1)  +
	\lambda_7(\Phi_2^{\dagger}\Phi_2)]\Phi_1^{\dagger}\Phi_2 + \textrm{h.c.}\big\}.
\end{align}
Here, the mass parameters $m_{11}^2$ and $m_{22}^2$ are real, $m_{12}^2$ is complex, the quartic couplings
$\lambda_{1...4}$ are real, and $\lambda_5$, $\lambda_6$, and $\lambda_7$ are in general complex. The two Higgs doublets
$\Phi_1$ and $\Phi_2$, both having hypercharge $Y=1$, can be decomposed into
\begin{align}
\Phi_1 = \begin{pmatrix} \phi^+_1 \\ \frac{1}{\sqrt{2}} (v_1 + \phi_1 + \text{i} a_1) \end{pmatrix}, \quad \Phi_2 =
\begin{pmatrix} \phi^+_2 \\ \frac{1}{\sqrt{2}} (v_2 + \phi_2 + \text{i} a_2) \end{pmatrix}
\end{align}
where $v_1$ and $v_2$ are the vacuum expectation values, $\phi^+_1$, $\phi^+_2$ two complex Higgs fields, and $\phi_1$,
$\phi_2$, $a_1$, $a_2$ the neutral Higgs fields.

In the matching conditions, we take into account loop-induced couplings of the``wrong" Higgs doublet to the
corresponding quarks, which renders the 2HDM a type III instead of the tree-level type II version, where one Higgs
doublet couples only to the up-type quarks and the other Higgs doublet couples to the down-type quarks and the charged
leptons. The Yukawa Lagrangian for the third generation is accordingly
\begin{align}\label{yukawas}
	\mathcal L_{\text{Yukawa}} = h_t' \epsilon_{ij} \Phi_1^i t_c Q^j  + h_t \epsilon_{ij} \Phi_2^i t_c Q^j - h_b
	\delta_{ij}\Phi_1^{*i} b_c Q^j - h_b'\delta_{ij}\Phi_2^{*i} b_c Q^j + \text{h.c.}\,.
\end{align}
Here, we follow the SUSY conventions and write all fields as left-handed fields. $Q$ is the quark doublet, $\epsilon_{12}
= 1$, and $t_c$ and $b_c$ are the left-handed top- and bottom-quark charge-conjugate fields, respectively. The Yukawa
couplings $h_t$ and  $h_b$ are the top- and  bottom-Yukawa couplings also present in the type II 2HDM case, while $h_t'$
and $h_b'$ denote the coupling to the ``wrong" Higgs doublet only existing in the type III case. We neglect
contributions from Yukawa couplings from the first two generations and neglct the $h_{\tau}$ and $h'_{\tau}$ Yukawa
couplings.  

The tree-level mass matrices, parameterized in terms of charged Higgs boson mass $M_{H^\pm}$, have the entries 
\begin{align}\nonumber
\mathcal{M}^{2}_{11} &= v^2 \left(c_{\beta}^2 \lambda_1
	+\frac{1}{2} s_{\beta}^2 \left[\lambda_4+\re(\lambda_5)\right] + 2 c_{\beta}s_{\beta} \re(\lambda_6)\right)+ s_{\beta}^2 M_{H^\pm}^2, \\\nonumber
\mathcal{M}^{2}_{12} &=  v^2 \left(c_{\beta} s_{\beta}
	\lambda_3   + \frac{1}{2} c_{\beta} s_{\beta}
	\left[\lambda_4+\re(\lambda_5)\right] +c_{\beta}^2 \re(\lambda_6) + s_{\beta}^2 \re(\lambda_7)\right)
	-c_{\beta}  s_{\beta}M_{H^\pm}^2,\\\nonumber
\mathcal{M}^{2}_{22} &= v^2 \left(s_{\beta}^2 \lambda_2 + \frac{1}{2} c_{\beta}^2  \left[\lambda_4+\re(\lambda_5)\right]+2 c_{\beta}s_{\beta}\re(\lambda_7) \right)
	+c_{\beta}^2 M_{H^\pm}^2, \\\nonumber
\mathcal{M}^{2}_{33} &= \frac{1}{2} s_{\beta}^2 \{v^2 \left[\lambda_4-\re(\lambda_5)\right]+2
	M_{H^\pm}^2\}, 
	\\\nonumber
\mathcal{M}^{2}_{34} &= -\frac{1}{2} c_{\beta} s_{\beta}  \{v^2 \left[\lambda_4-\re(\lambda_5)\right]+2
	M_{H^\pm}^2\},\\\nonumber
\mathcal{M}^{2}_{44} &= \frac{1}{2} c_{\beta}^2  \{v^2 \left[\lambda_4-\re(\lambda_5)\right]+2
	M_{H^\pm}^2\},\\\nonumber
\mathcal{M}^{2}_{13} &= \frac{1}{2} s_{\beta} v^2 \left[s_{\beta} \im(\lambda_5) + 2 c_{\beta} \im(\lambda_6)\right] = -{\tan \beta} \mathcal{M}^{2}_{14}, \\
\mathcal{M}^{2}_{23} &= \frac{1}{2} s_{\beta} v^2 \left[c_{\beta} \im(\lambda_5)+2 s_{\beta} \im(\lambda_7)\right] =-{\tan \beta}\mathcal{M}^{2}_{24}, \label{eq:2HDMmassmatrixneutral}\\\nonumber
\mathcal{M}^{2}_{+_{11}} &= s_{\beta}^2 M_{H^\pm}^2, \\\nonumber
\mathcal{M}^{2}_{+_{12}} &= -c_{\beta} s_{\beta}M_{H^\pm}^2  = \mathcal{M}^{2}_{+_{21}},\\
\mathcal{M}^{2}_{+_{22}} &= c_{\beta}^2 M_{H^\pm}^2  \label{eq:2HDMmassmatrixcharged}
\end{align}
\be\nonumber
\text{with \quad} v^2\equiv \sqrt{v_1^2+v_2^2},
\ee
where $\mathcal{M}^2$ is the neutral Higgs mass matrix in the $(\phi_1, \phi_2,
a_1, a_2)$ basis and $\mathcal{M}^{2+}$ is the charged Higgs mass matrix  in
the $(\phi_1^+, \phi_2^+)$ basis.


\section{Matching the MSSM to the 2HDM}
\label{se:matching}
\paragraph{}
The complex MSSM Higgs potential of Eq.~\eqref{higgs_potential} is matched to the general type-III 2HDM given above at the
one-loop level at the scale $M_{\text{s}}$. The doublets $H_u$ and $H_d$ from Eq.~\eqref{higgs_potential} are related to those
of Eq.~\eqref{THDM_potential} by 
\be
\Phi_1\equiv-i\sigma_2H_d^*, \qquad \Phi_2 \equiv H_u.
\ee
At tree level, the matching conditions for the quartic Higgs couplings are

\begin{align}\nonumber
\lambda_1 &= \frac{1}{4}(g^2+g_y^2),\\\nonumber
\lambda_2 &= \frac{1}{4}(g^2+g_y^2),\\\nonumber
\lambda_3 &= \frac{1}{4}(g^2-g_y^2),\\\nonumber
\lambda_4 &= -\frac{1}{2}g^2,\\
\lambda_5 &= \lambda_6 = \lambda_7 = 0. 
\end{align}

For the Yukawa couplings, one obtains 
\begin{equation}
h_t^{\mathrm{2HDM}} = h_t^{\mathrm{MSSM}},\qquad h_b^{\mathrm{2HDM}} = h_b^{\mathrm{MSSM}},\qquad h_t'^{\mathrm{2HDM}}= h_b'^{\mathrm{2HDM}}=0.
\end{equation}
Here we are assuming that we know all the parameters of the MSSM and match the MSSM to the 2HDM. This means all
couplings given below are assumed to be MSSM couplings unless we explicitly state otherwise and we drop the superscript ${}^{\mathrm{MSSM}}$.

The one-loop threshold corrections are calculated under three assumptions. Firstly, all supersymmetric soft-breaking
mass parameters are assumed to share the common mass scale $M_{\text{s}}$, in particular $M_{\text{s}}= M_{L_Q} =
M_{R_U} = M_{R_D} = M_3$.  Secondly, all Yukawa couplings are assumed to vanish except the top and bottom Yukawa couplings,
and only contributions proportional to powers of these Yukawa couplings or the strong gauge coupling are included. This
amounts to only including third-generation squarks in loops when deriving the thresholds and neglecting terms of
$\mathcal O(g^a g_y^b)$ with $a+b = 4$. Lastly, all loop functions are evaluated in the limit of zero external momenta.
The diagrams were evaluated using \texttt{FeynArts} and \texttt{FormCalc} \cite{Hahn:2000kx},\cite{Hahn:1998yk}, and the
loop functions are evalued using \texttt{ANT} \cite{Angel:2013hla}. 

The results given here agree with the complex results given in Ref.~\cite{Carena:2015uoe} up to gauge contributions to the
couplings $\lambda_6$ and $\lambda_7$, which, however, are found in the work of  Ref.~\cite{Haber:1993an} and
 Ref.~\cite{Bahl:2018jom} in the real case. 
\begin{figure}
	\centering
	\begin{subfigure}[]{.4\textwidth}
		\centering	\includegraphics[width=0.7\textwidth]{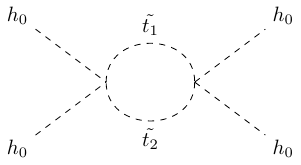}
		\caption{box diagram with squarks}
		\label{fig:boxes}
	\end{subfigure}\qquad
	\begin{subfigure}[]{.4\textwidth}
			\centering\includegraphics[width=0.7\textwidth]{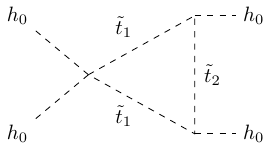}
		\caption{triangle diagram with squarks}
		\label{fig:triangles}
	\end{subfigure}
	\caption{Two sample diagrams contributing to the one-loop threshold of the quartic Higgs couplings. }
	\label{fig:diagrams}
\end{figure}

Box diagrams like those of Fig.~\ref{fig:boxes} lead to the following corrections to the quartics   
\begin{align}
\Delta\lambda_{1}^{(4)} &= -\frac{\kappa}{2 } \Big\{|\hat{A}_b|^4 h_b^4+h_t^4 |\hat{\mu}|^4\Big\},\\
\Delta\lambda_{2}^{(4)} &= -\frac{\kappa}{2 } \Big\{|\hat{A}_t|^4 h_t^4+h_b^4 |\hat{\mu}|^4\Big\},\\\nonumber
\Delta\lambda_{3}^{(4)} &= \frac{\kappa}{2 } \Big\{-|\hat{A}_b|^2 |\hat{A}_t|^2 h_b^2 h_t^2-|\hat{A}_b|^2 h_b^4
	|\hat{\mu}|^2-|\hat{A}_t|^2 h_t^4|\hat{\mu}|^2-h_b^2 h_t^2 |\hat{\mu}|^4 \\&\quad+(\hat{A}_b \hat{A}_t^* +
	\hat{A}_b^* \hat{A}_t) h_b^2 h_t^2 |\hat{\mu}|^2\Big\},\\\nonumber
\Delta\lambda_{4}^{(4)} &= \frac{\kappa}{2 } \Big\{|\hat{A}_b|^2 |\hat{A}_t|^2 h_b^2 h_t^2-|\hat{A}_b|^2 h_b^4
	|\hat{\mu}|^2-|\hat{A}_t|^2 h_t^4|\hat{\mu}|^2+h_b^2 h_t^2 |\hat{\mu}|^4 \\&\quad- (\hat{A}_b \hat{A}_t^* +
	\hat{A}_b^* \hat{A}_t)  h_b^2 h_t^2 |\hat{\mu}|^2 \Big\},\\
\Delta\lambda_{5}^{(4)} &= 
-\frac{\hat{\mu}^2}{2 }\kappa\Big\{\hat{A}_b^2 h_b^4 
+\hat{A}_t^2 h_t^4\Big\},\\
\Delta\lambda_{6}^{(4)} &= \frac{\hat{\mu}}{2 } \kappa\Big\{|\hat{A}_b|^2 \hat{A}_b  h_b^4 
	+ \hat{A}_t h_t^4  |\hat{\mu}|^2\Big\},\\
\Delta\lambda_{7}^{(4)} &= \frac{\hat{\mu}}{2 } \kappa\Big\{|\hat{A}_t|^2 \hat{A}_t h_t^4 
	+\hat{A}_b h_b^4   |\hat{\mu}|^2\Big\},\\
\kappa&\equiv\frac{1}{16\pi^2}.
\end{align}
All hatted parameters above and following in the rest of the paper are normalized to the scale $M_\text{s}.$

The triangle diagrams like those of Fig.~\ref{fig:triangles} give 
\begin{align}
\Delta\lambda_{1}^{(3)} &= \frac{3}{4 }\kappa \Big\{-|\hat{A}_b|^2 h_b^2 \Big(g^2+g_y^2-8h_b^2\Big)+\Big(g^2+g_y^2\Big)
	h_t^2 |\hat{\mu}|^2\Big\},\\
\Delta\lambda_{2}^{(3)} &= \frac{3}{4 }\kappa \Big\{-|\hat{A}_t|^2 h_t^2 \Big(g^2+g_y^2-8h_t^2\Big)+\Big(g^2+g_y^2\Big)
	h_b^2 |\hat{\mu}|^2\Big\},\\\nonumber
\Delta\lambda_{3}^{(3)} &= 
-\frac{3}{8}\kappa \Big\{h_t^2|\hat{A}_t|^2 (g^2-g_y^2 -4  h_b^2)
+h_b^2\ |\hat{A}_b|^2 \Big(g^2-g_y^2-4 h_t^2\Big)\\\nonumber &\quad-h_b^2 |\hat{\mu}|^2\left[g^2-g_y^2+4 (h_b^2 -h_t^2)\right]
	 -h_t^2 |\hat{\mu}|^2\left[g^2-g_y^2 + 4( h_t^2 - h_b^2)\right]  \\	
	&\quad - 4 h_b^2 h_t^2(\hat{A}_b \hat{A}_t^* + \hat{A}_b^* \hat{A}_t) \Big\},\\\nonumber
\Delta\lambda_{4}^{(3)} &= \frac{3}{4 }\kappa \Big\{+h_t^2 |\hat{A}_t|^2 \left(g^2 - 2 h_b^2\right) +h_b^2 |\hat{A_b}|^2  \left(g^2- 2 h_t^2\right)
 \\&\quad - |\hat{\mu}|^2(h_t^2+h_b^2)\left[g^2- 2 (h_b^2 +h_t^2)\right] -2 h_b^2 h_t^2 (\hat{A_b} \hat{A}_t^* + \hat{A}_b^*
	\hat{A}_t)  \Big\},\\
\Delta\lambda_{5}^{(3)} &= 0,\\
\Delta\lambda_{6}^{(3)} &= \frac{3 \hat{\mu}}{8 }\kappa \Big\{\hat{A_b} h_b^2  \Big(g^2+g_y^2-8 h_b^2\Big)
	- \hat{A}_t h_t^2  \Big(g^2+g_y^2\Big)\Big\},\\
\Delta\lambda_{7}^{(3)} &= -\frac{3 \hat{\mu}}{8 }\kappa \Big\{\hat{A_b} \Big(g^2+g_y^2\Big) h_b^2
	-\hat{A}_t  h_t^2 \Big(g^2+g_y^2-8 h_t^2\Big)\Big\}.
\end{align}
There are also contributions coming from the redefinition of the Higgs doublets. Squark loops induce
mixing between the scalar fields, which must be accounted for in order to preserve canonically normalized kinetic terms
for the scalar fields in the Lagrangian. This is done by redefining the Higgs doublet fields in the following manner

\be\label{doublet_redefine}
	\begin{pmatrix}\Phi_1 \\ \Phi_2 \end{pmatrix} \rightarrow \begin{pmatrix}\Phi_1 \\ \Phi_2 \end{pmatrix} -
\frac{1}{2}\begin{pmatrix}
\Delta Z_{\Phi_1 \Phi_1} & \Delta Z_{\Phi_1 \Phi_2} \\
\Delta Z_{\Phi_2 \Phi_1} & \Delta Z_{\Phi_2 \Phi_2}
 \end{pmatrix}\begin{pmatrix}\Phi_1 \\ \Phi_2 \end{pmatrix}. 
\ee
The SU(2) invariance ensures that the corrections can be applied to the complete Higgs doublets and not only the component fields.
The expressions for the wave-function-correction factors $\Delta Z_{\Phi_i \Phi_j}$ can be derived via the finite parts of the derivatives of the self energies in the electroweak interaction basis $\Sigma'_{\phi_i \phi_j}$ with $\phi_{\{i, j\}} = \{\phi_1, \phi_2, a_1, a_2\}$, corresponding to a $\overline{\mathrm{MS}}$ renormalized self energy,
\begin{align}
\Delta Z_{\Phi_1 \Phi_1}  &= \frac{1}{2}\left(\Sigma'_{\phi_1 \phi_1} + \Sigma'_{a_1 a_1} \right)= \kappa\frac{h_t^2 |\hat{\mu}|^2 + h_b^2 |\hat{A_b}|^2}{2}, \\
\Delta Z_{\Phi_1 \Phi_2}  &= \frac{1}{2}\left( \Sigma'_{\phi_1 \phi_2} + \text{i} \Sigma'_{\phi_1 a_2} -  \text{i} \Sigma'_{\phi_2 a_1} + \Sigma'_{a_1 a_2} \right) =-\kappa\frac{\hat{\mu} (h_t^2 \hat{A}_t +
	h_b^2 \hat{A_b})}{2 }, \\
\Delta Z_{\Phi_2 \Phi_1}  &=  \Delta Z_{\Phi_1 \Phi_2}^*, \\
\Delta Z_{\Phi_2 \Phi_2}  &= \frac{1}{2}\left( \Sigma'_{\phi_2 \phi_2} + \Sigma'_{a_2 a_2} \right)=   \kappa\frac{h_t^2 |\hat{A}_t|^2 + h_b^2 |\hat{\mu}|^2}{2}.
\end{align}
In Ref.~\cite{Bahl:2018jom}, it was shown for the CP-even Higgs boson fields that this choice for the wave-function-correction factors together with an appropriate choice of correction of the mixing angle leads to the physical fields being the same in the MSSM and the 2HDM at the matching scale as required.   The field redefinitions lead to
the following threshold corrections  
\begin{align}
&\Delta\lambda_{1}^{(2)} = -\frac{g^2+g_y^2}{4 } 
 \kappa\Big(h_b^2 |\hat{A_b}|^2
+ h_t^2 |\hat{\mu}|^2\Big),
\\
&\Delta\lambda_{2}^{(2)} = -\frac{g^2+g_y^2}{4 } 
 \kappa\Big(h_t^2 |\hat{A}_t|^2 
+ h_b^2|\hat{\mu}|^2\Big)),
\\
&\Delta\lambda_{3}^{(2)} = -\frac{g^2-g_y^2}{8 } 
\kappa\Big( h_t^2 \left(|\hat{A}_t|^2 +|\hat{\mu}|^2\right) 
+ h_b^2 \left(|\hat{A_b}|^2+|\hat{\mu}|^2\right) \Big),
\\
&\Delta\lambda_{4}^{(2)} = \frac{g^2\kappa}{4}\Big(h_b^2(|\hat{A_b}^2| + |\hat{\mu}|^2) + h_t^2(|\hat{A}_t^2| +
	|\hat{\mu}|^2)\Big),\\
&\Delta\lambda_{5}^{(2)} = \Delta\lambda_{6}^{(2)} = \Delta\lambda_{7}^{(2)} = 0.
\end{align}

Finally, the Yukawa couplings receive the one-loop corrections resulting in the following 2HDM Yukawa couplings at the matching scale (including the tree-level contribution): 
\begin{align} 
h_t^{\text{2HDM}} &= h_t\bigg\{1 -
\kappa\bigg[\frac{4}{3}g_s^2\left(\hat{A}_t\hat{M}^*_3
 - 1\right) +
 \frac{1}{4}\left(h_t^2  |\hat{A}_t|^2 + h_b^2 |\hat{\mu}|^2\right)
 \nonumber \\& \qquad
{{- h_b^2 |\hat{\mu}|^2 F_3(|\hat{\mu}|^2)  -  \frac{1}{8} (h_b^2 + 3 h_t^2)F_1(|\hat{\mu}|^2)}}
 \bigg]\bigg\},\label{eq:htthres} 
\\ 
h_b^{\text{2HDM}} &= h_b\bigg\{1 -
\kappa\bigg[\frac{4}{3}g_s^2\left(\hat{A_b}\hat{M}^*_3
 - 1\right) +
	\frac{1}{4}\left(h_b^2  |\hat{A_b}|^2 + h_t^2 |\hat{\mu}|^2\right)  \nonumber \\& \qquad
{{- h_t^2 |\hat{\mu}|^2 F_3(|\hat{\mu}|^2)  -  \frac{1}{8} (h_t^2 + 3 h_b^2)F_1(|\hat{\mu}|^2)}}
 \bigg]\bigg\},\label{eq:hbthres} 
\\
h_t^{'\text{2HDM}} &= \kappa h_t\left\{\frac{4}{3}g_s^2\hat{\mu}^*\hat{M}^*_3 + \frac{1}{4}
	\left(h_b^2 \hat{A}_b^* \hat{\mu}^*  +  h_t^2 \hat{A}_t^*\hat{\mu}^*\right) {{ - h_b^2 \hat{A}_b^* \hat{\mu}^* F_3(|\hat{\mu}|^2)}} \right\},
\\
h_b^{'\text{2HDM}} &= \kappa h_b\left\{\frac{4}{3}g_s^2\hat{\mu}^*\hat{M}^*_3 + \frac{1}{4}
	\left(h_b^2 \hat{A}_b^* \hat{\mu}^*  + h_t^2 \hat{A}_t^*\hat{\mu}^*\right) {{ - h_t^2 \hat{A}_t^* \hat{\mu}^* F_3(|\hat{\mu}|^2)}}\right\}
\end{align}
with
    \begin{alignat}{3}
      F_1(x) &=  -\frac{1 - 4 x +  x^2 \left[3 - 2 \ln(x)\right]}{(1 - x)^2} , \qquad & F_1(1) &= 0, \\
      F_3(x) &=  -\frac{1 - x \left[1 -  \ln(x)\right]}{(1 - x)^2},  & F_3(1) &= \frac{1}{2}.
\end{alignat}
It should be noted that, since the absolute value of the gluino mass parameter is $|M_3| = M_S$, $\hat{M}_3$ is
just a phase factor $\hat{M}_3= \text{e}^{\text{i} \varphi_{M_3}}$ where $\varphi_{M_3}$ is the phase of the gluino
mass parameter. 

We do not calculate threshold corrections to parameters such as $\tan \beta$, since they do not enter in the MSSM
threshold corrections and, hence, are only needed as 2HDM parameters.


\section{Calculating the Higgs-mass Spectrum and the Mixing}\label{mass_calculation}
\label{se:mass}
\paragraph{}
The Higgs masses are determined completely once all the parameters of the MSSM
at the scale $M_s$ are given. These are the necessary boundary conditions for
solving the RGEs and obtaining the values for the relevant couplings at the
scale where the masses are calculated. However, not all the relevant input
parameters are given at the same scale. The soft parameters of the MSSM are
given as user-defined input at the scale $M_s$ (except for $\tan\beta$, which is
defined  at the scale $M_{H^+}$, and, hence as a 2HDM parameter), while all SM
couplings relevant to the calculation are fixed at the electroweak scale.  

There are different ways to approach this mixed-scale boundary-value issue. The
``bottom up" approach starts with the low-energy scale values from the SM and
evolves the parameters up to the high-energy scale taking matching effects into
account on the way up to the high-energy scale and guessing the values of the
first parameters such as $\lambda_i$. In an iterative procedure, evolving the
parameters up and down, the complete set of parameters at a single energy scale
is found.  With the ``top down'' approach, which has been exploited also in
Refs.~\cite{Bahl:2016brp, Bahl:2018jom}, one guesses inital values for the high
scale MSSM parameters and evolves all the couplings down to $M_t$. Here, the
couplings calculated from the EFT procedure are compared to the experimentally
fixed values, and the high scale parameters are adjusted to minimize the
differences using a numerical algorithm. This way, evolving
parameters up to the high scale can be avoided. We adapt the ``top down''
approach, employing an upwards evolution of the parameters just for the initial
guess. The process is sketched in Fig.~\ref{fig:process} and the single steps
are described in the following:
\begin{enumerate}
\item First, initial values for the high scale MSSM couplings as a first guess have to be found To obtain these,
\begin{enumerate}
 \item we start at the scale $M_t$ (with $M_t$ being the top \textit{pole
mass}) where the SM couplings are fixed, guess a value for the SM quartic
Higgs coupling of $\lambda_{\text{SM}} = 0.25$, and evolve the SM couplings up to
the intermediate scale $M_{H^+}$ using SM RGEs obtained from {\tt
Sarah}~\cite{Staub:2012pb, Staub:2013tta}.
\item At the scale $M_{H^+}$, it is assumed that all the 2HDM quartic couplings $\lambda_1$,\dots, $\lambda_7$, ``wrong''
Yukawa couplings $h_t'$ and $h_b'$, and the phases of the Yukawa couplings $\varphi_{h_t}$, $\varphi_{h_t'}$, $\varphi_{h_b}$, and $\varphi_{h_b'}$ are zero, and the 2HDM Yukawa couplings $h_t$ and $h_b$ are calculated accordingly via the tree-level matching of the Yukawa couplings,
\be\label{eq:top}
	h_t^{\mrm{2HDM}} = \frac{1}{\sin\beta}\, y_t^{\mrm{SM}}, 
\ee
\be\label{eq:bottom}
	h_b^{\mrm{2HDM}} = \frac{1}{\cos\beta}\, y_b^{\mrm{SM}}. 
\ee
Then, the 2HDM couplings are evolved up to the scale $M_s$ using the full two-loop 2HDM RGEs
including complex phases. For the gauge, Yukawa, and quartic couplings, we have calculated these implementing the general prescription first developed by
Refs.~\cite{Vaughn:1983,Vaughn:1984,Vaughn:1985,Luo:2002ti} and expanded upon in Ref.~\cite{Schienbein:2018fsw} to account for
kinetic mixing of scalar fields in the presence of multiple Higgs doublets. For the running vevs, we use the formulae
from Refs.~\cite{Sperling:2013eva,Sperling:2013xqa}. We have checked our results for the couplings with the findings of the authors of Refs.~\cite{Oredsson:2018yho, Thomsen:2021ncy}
and  find agreement\footnote{Thanks to the work in Refs.~\cite{Schienbein:2018fsw} and \cite{Thomsen:2021ncy}, we became aware of a typo in Ref.~\cite{Vaughn:1984} in Eq.~(3.3) that, at first, also entered into our RGEs. This first version is in agreement with Ref.~\cite{Oredsson:2018yho} while the current version agrees with Ref.~\cite{Thomsen:2021ncy}}. 
\item As our first guess, the values of the  2HDM
gauge and Yukawa couplings emerging from the previous step are taken to determine the initial values of the MSSM gauge and Yukawa couplings,
\begin{align}
c^{\text{MSSM}} = c^{\text{2HDM}} \quad \text{with} \quad  c = g_y, g, g_s, h_t, h_b. 
\end{align}
It should be noted that
in the MSSM, the ``wrong'' Yukawa couplings are purely loop-induced and that the
Yukawa phases can be absorbed into the fields. Hence, we only have the real parameters
$h_t^\mrm{MSSM}$ and $h_b^\mrm{MSSM}$.
\end{enumerate}
\item Now, the MSSM parameters are given by the gauge and Yukawa couplings obtained in step~1 (or adapted in the minimization procedure) and the soft SUSY breaking parameters $A_t$, $A_b$, $\varphi_{M_3}$ as well as the parameter $\mu$ defined as input at the scale $M_s$ used in the following steps:
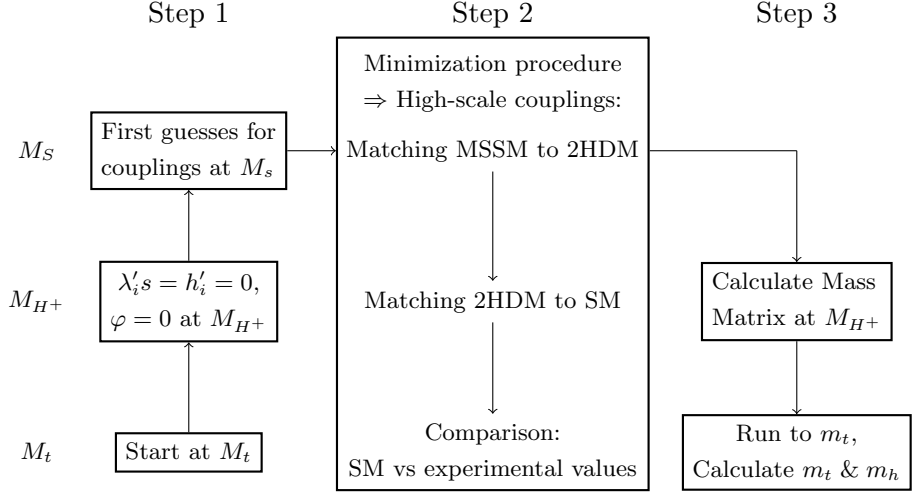
\begin{figure}
\centering
	\begin{tikzpicture}
		\node (a) at (-2,0) {\footnotesize{$M_t$}};
		\node (b) at (-2,2) {\footnotesize{$M_{H^+}$}};
		\node (c) at (-2,4) {\footnotesize{$M_S$}};
		\node (A) at (0,0) [draw,thick] {\footnotesize{Start at $M_t$}};
		\node (B) at (0,2) [draw,thick,align=center] {\footnotesize{$\lambda_i's = h^\prime_i=0,$} \\
		\footnotesize{$\varphi=0$ at $M_{H^+}$}};
		\node (C) at (0,4) [draw,thick,align=left] {\footnotesize{First guesses for} \\ \footnotesize{couplings at $M_s$}};
		\node       at (0,5.8) {Step 1};
		\node       at (4,5.8) {Step 2};
		\node       at (8,5.8) {Step 3};
		\draw (A) [->] -- (B);
		\draw (B) [->] -- (C);
		\node (D2) at (4,4.9) [
		align=center]{\footnotesize{Minimization procedure} \\ \footnotesize{$\Rightarrow$ High-scale couplings:}};
		\node (D) at (4,4)[align=center]{\footnotesize{Matching MSSM to 2HDM}};
		\node (Ee) at (4, 2)[align=center]{\footnotesize{Matching 2HDM to SM}};
		\node (Ff) at (4, 0)[align=center]{\footnotesize{Comparison:}\\\footnotesize{SM vs experimental values}};
		\draw[thick] (1.95,-0.5) rectangle (6.05,5.5);
		\node (E) at (8.,2) [draw,thick,align=left]{\footnotesize{Calculate Mass} \\ \footnotesize{Matrix at $M_{H^+}$}};
		\node (F) at (8,0) [draw,thick,align=center]{\footnotesize{Run to $m_t$,} \\ \footnotesize{Calculate $m_t\;\mrm{\&}\;m_h$}};
		\draw (D) [->] -- (Ee);
		\draw (Ee) [->] -- (Ff);
		\draw (C) [->] -- (D);
		\draw (D) [-] -- (8,4);
		\draw (8,4)[->] -- (E);
		\draw (E) [->] -- (F);
	\end{tikzpicture}
\caption{Pictorial description of the mass calculation.}
\label{fig:process}
\end{figure}
\begin{enumerate} 
\item  With the MSSM parameters, the 2HDM
couplings are calculated using the matching conditions given in Sect.~\ref{se:matching}. These MSSM threshold corrections give the non-vanishing values for the 2HDM quartic couplings $\lambda_1, \dots, \lambda_7$,  
the ``wrong'' Yukawa couplings $h_t^{\prime\mrm{2HDM}}$, $h_b^{\prime\mrm{2HDM}}$, and  the Yukawa phases of the 2HDM $\varphi_{h_t}$, $\varphi_{h_t'}$, $\varphi_{h_b}$, and $\varphi_{h_b'}$.  The couplings are then run down to the scale $M_{H^+}$.
\item \label{SM_THDM_matching} Then, the 2HDM
is matched to the SM. The tree-level matching conditions for the SM Yukawa couplings $y_t$ and $y_b$ to the 2HDM ones $h_t$, $h_b$, $h_t'$ and $h_b'$ are 
\begin{align}\nonumber
	&\sqrt{|h_t|^2\sin^2\beta + 2|h_t||h_t^\prime|\cos\beta\sin\beta\cos(\varphi_{h_t} - \varphi_{h_t^\prime}) +
	|h_t^\prime|^2\cos^2\beta} = y_t\\\label{eq:quark_mass_match}
	&\sqrt{|h_b|^2\cos^2\beta + 2|h_b||h_b^\prime|\cos\beta\sin\beta\cos(\varphi_{h_b} - \varphi_{h_b^\prime}) +
	|h_b^\prime|^2\sin^2\beta} = y_b
\end{align}
where $\varphi_\chi$ is the phase of the coupling $\chi$. The quartic Higgs coupling in the SM
$\lambda^{\mrm{SM}}$ can be calculated at tree-level via
\begin{equation}\label{eq:lamSM2HDM}
\lambda_{\text{SM}} = c_\beta^4 \lambda_1 + 4 c_\beta^3 s_\beta \re(\lambda_6) + 
2 c_\beta^2 s_\beta^2\left[\lambda_3 + \lambda_4 + \re(\lambda_5)\right] + 4 c_\beta s_\beta^3 \re(\lambda_7) + s_\beta^4
\lambda_2. 
\end{equation}
The one-loop threshold correction to $\lambda_{SM}$ is obtained by integrating out the heavy Higgs bosons. In the real case
where all phases are set to zero, the answer is known in closed form \cite{Bahl:2018jom}. In the complex case, on the
other hand, the calculation is complicated by the $4\times4$ neutral mixing and mass matrices. We have
evaluated the full threshold corrections numerically for the complex case, which leads to the problem that the result includes contributions of order 
$\mathcal O(v/M_{H^+})$ that are ignored elsewhere in the calculation. Comparing the results for the Higgs masses using only the tree-level matching, the one-loop threshold of Ref.~\cite{Bahl:2018jom}, and the full one-loop threshold including  $\mathcal O(v/M_{H^+})$ terms leads to very small differences, so we can neglect the one-loop threshold entirely. Similarly, the one-loop 2HDM threshold corrections to the SM Yukawa couplings are numerically negligible. 
\item \label{SM_evolution}  In the next step, the SM couplings are evolved from the scale $M_{H^+}$ down to $M_t$ and checked against the
  experimental values for\footnote{Including a check of the value of the vev does not change the result within our numerical accuracy.} $g_y$, $g$, $g_s$, $y_t$, $y_b$. We repeat this procedure,
  adjusting the high scale MSSM couplings each time to minimize the differences between the SM couplings at the low-scale and the experimental values until
  good agreement is found. 
  \end{enumerate}
 \item \label{masscalculation} Via the minimization procedure in step~2, we obtained a final set of value for all MSSM high scale parameters. These are
   evolved down one last time to $M_{H^+}$. At this stage, all
the low-scale 2HDM parameters necessary for computing the Higgs masses are determined, and one could in principle
calculate the eigenvalues of the loop-corrected mass matrix at the scale
$M_{H^+}$ and determine the pole masses. However, this will lead to terms
containing potentially large logarithms of $\ln(M_{H^+}/m_t)$, which are
additionally enhanced by factors of the large top-Yukawa coupling. These terms
originate  from the one-loop corrections in the conversion of the
$\overline{\text{MS}}$ mass to the pole mass. Therefore, we considered three
conceptionally different methods to calculate the Higgs-boson masses: In the
case that the charged Higgs boson is sufficiently light, the 2HDM can be used as
the low-energy theory (options (a) and (b) below). If the charged Higgs boson is
heavy, then the SM is the appropriate low-energy theory and a matching procedure for the 2HDM and the SM is performed at the scale $M_{H^+}$ (option (c)). Finally, we apply an approximation that interpolates between both results (option (d)). In the following, we list the options and include some details about the calculation:
\begin{enumerate}
\item \label{2HDMatMHp} The parameters are taken at the scale $\mu_{\text{ren}} = M_{H^+}$ and the on-shell Higgs masses are calculated via the zeros of the determinant
\begin{align}\label{eq:det}
\det\left[p^2 - \mathcal M^2 (\mu_{\text{ren}}) + \hat{\Sigma}(\mu_{\text{ren}}, p^2) - \hat{T}\right] = 0
\end{align}
expanded up to one-loop order
where $\hat{\Sigma}(\mu_{\text{ren}}, p^2)$ denotes the top and bottom Yukawa contributions to the self energy matrix in
the $\overline{\text{MS}}$ renormalization scheme at momentum $p^2$. To ensure the proper minimum
of the effective potential, tadpole contributions $\hat{T}$  originating from top and
bottom loops have to be taken into account. The entries of the matrix $\hat{T}$ are given in the appendix~\ref{tadpole} in terms of tadpole contributions in the interaction basis. The mass matrix $\mathcal M^2
(\mu_{\text{ren}})$ has the form of the tree-level mass matrix given in
Eqs.~\eqref{eq:2HDMmassmatrixneutral} with the parameters evaluated at the scale
$\mu_{\text{ren}} = M_{H^+}$, where the charged Higgs mass is given in the
$\overline{\text{MS}}$ scheme.
\item \label{2HDMatmt} In this option, the low-energy theory is still the 2HDM, however, the
parameters are evolved down to the \textit{running} top-quark mass $m_t$
calculated in terms of 2HDM parameters, and Eq.\eqref{eq:det} is evaluated at
the scale $\mu_{\text{ren}} = m_t$ where the $\overline{\text{MS}}$ mass of the charged Higgs boson $M_{H^+}$ is interpreted as given at the scale\footnote{Within the calculation, it is consistent to use $M_{H^+}(m_t)$ instead of $M_{H^+}(M_{H^+})$ as an input---$M_{H^+}(M_{H^+})$ is chosen  in step~\ref{2HDMatMHp} as input. However, when comparing both approaches of step~\ref{2HDMatMHp} and of step~\ref{2HDMatmt}, one has to be careful with the interpretation of the results. We find that the relative difference between $M_{H^+}(m_t)$  and $M_{H^+}(M_{H^+})$ is at the per-mille level in the parameter region where the 2HDM calculation is applicable and we ignore this difference.}
 $m_t$. Using this scale choice, the logarithms
$\ln(\mu_{\text{ren}}/m_t)$ in the self energies vanish, since we evaluate the
self-energies using the running top and bottom masses. This method is only valid
as long as $M_{H^+}$ is not much larger than $m_t$ since the 2HDM RGEs are not
the correct RGEs for evolving the couplings below the scale $M_{H^+}.$
\item 
  In this method, the SM is decoupled completely from the 2HDM and treated as
the low-energy theory. This method applies when $M_{H^+}\gg m_t$. In this case,
the heavy Higgs bosons of the 2HDM are decoupled by matching the 2HDM to the SM
in the same way as step \ref{SM_THDM_matching}, and the SM couplings are
evolved down to $m_t$. In this case, however, $m_t$ is calculated in terms of SM
parameters only. The $\overline{\text{MS}}$ mass of the lightest Higgs boson is
taken to be $v^2\lambda_{\mathrm{SM}}(m_t)$, which is converted to the pole mass
via 
\begin{align}\label{eq:polemass}
p^2 - v^2\lambda_{\mathrm{SM}}(m_t) + \hat{\Sigma}^\text{SM} - \hat{T}^\text{SM} = 0
\end{align}
where $ \hat{\Sigma}^\text{SM}$ and $\hat{T}^\text{SM}$ are the self-energy and
tadpole contributions of the SM-like Higgs boson of $\mathcal O(\alpha_t)$ and $\mathcal O(\alpha_b)$ with $\alpha_{\{t,b\}} = y_{\{t,b\}}^2/(4\pi)$ evaluated in the
$\overline{\text{MS}}$ scheme. In this option, since the heavy 2HDM Higgs
bosons are decoupled, all information about the heavy Higgs bosons at the
scale $m_t$ is encoded in the size of the couplings and their masses can be
estimated to be of order $M_{H^+}$. However, at the scale $M_{H^+}$, one can
still obtain direct information about the heavy Higgs bosons.
\item Finally, in this option, an approximation is exploited that allows one to
resum large logarithms in the scenario of $M_{H^+}\gg m_t$ while still retaining
information about the heavy Higgs bosons and the mixing between them and the
lightest Higgs boson. In order to do so, firstly, we only consider top-Yukawa
effects. We do not attempt to resum logarithms proportional to other couplings
in the 2HDM. This
approximation is good in the low-$\tan\beta$ regime where $h_b^{\mathrm{2HDM}}$
is small, which is also the most phenomenologically relevant regime, especially
for low values of the mass of the charged Higgs boson. The second assumption is
that for the resummation of these logs, our 2HDM can be considered a classic
type-II CP-even 2HDM where only one Higgs doublet couples to the top quarks.
This is because the logarithms we wish to resum arise from one-loop corrections,
and CP-violating and ``wrong-type" Yukawa couplings are already loop-suppressed,
so any effect these couplings may have will be suppressed by an extra loop
factor.

With this regime in mind, we wish to incorporate the effect of running from
$M_{H^+}$ to $m_t$ (with $m_t$ evaluated with parameters of the 2HDM) into the full 2HDM neutral mass matrix at the scale
$M_{H^+}$. To begin, we evaluate $\lambda_{\text{SM}}$ and
$v^{\overline{\mrm{MS}}}$ at $M_{H^+}$ and $m_t$ according to
steps~\ref{SM_THDM_matching} and \ref{SM_evolution}, but we take $y_b = 0$ into
account when running the SM couplings down to $m_t$. Then, at the scale
$M_{H^+}$, we rotate into the so-called ``Higgs Basis'' \cite{Branco:1999fs,
Gunion:2002zf, Haber:2015pua}, defined by
\begin{gather}\label{eq:higgs_basis}
H_1 = c_{\beta}\Phi_1 + s_{\beta}\Phi_2, \qquad H_2 = c_{\beta}\Phi_2 - s_{\beta}\Phi_1,\\\nonumber
H_1 = \bpm h_1^+\\\frac{1}{\sqrt{2}}(v+h_1+ib_1)\epm,\qquad H_2 = \bpm h_2^+\\\frac{1}{\sqrt{2}}(h_2
	+ib_2)\epm\\\nonumber
s_\beta\equiv \sin\beta, \qquad c_\beta\equiv\cos\beta,\qquad \tan\beta \equiv\frac{v_2}{v_1},\qquad v^2\equiv v_1^2 + v_2^2,
\end{gather}
where $ h_j^+$, $h_j$, and $b_j$ with $j =1, 2$ are the charged, the
CP-even, the CP-odd Higgs fields in the Higgs basis, respectively. In this
basis, only Higgs doublet $H_1$ gets a vev $v$ and can therefore be identified
with the SM Higgs doublet. The mass matrix in the Higgs basis can be obtained
via $\mathcal M^\text{Higgs} = \mathcal U \mathcal M^2 \mathcal U^\dagger$, where
$\mathcal M^2$ is given in Eq.~\eqref{eq:2HDMmassmatrixneutral} and
\begin{align}\label{eq:Umatrix}
\mathcal U = \begin{pmatrix} U &0 \\ 0 &U\end{pmatrix} \quad \text{with} \quad U = \begin{pmatrix} c_\beta &s_\beta \\ -s_\beta & c_\beta \end{pmatrix}, 
\end{align} 
and the (1,1) component of $\mathcal M^\text{Higgs}$ can be identified with
$v^{\mrm{SM}}\lambda^{\mrm{SM}}$ leading to the threshold condition given
in Eq.~\eqref{eq:lamSM2HDM}. The submatrix of $\mathcal M^\text{Higgs}$ given by
the second and third row and column describe the heavy Higgs bosons.

Before continuing, we note two relevant facts. First, in the decoupling limit
\cite{Gunion:2002zf}\footnote{Originally, the decoupling limit was formulated
for $M_A \gg v$ where $M_A$ is the mass of the CP-odd Higgs boson in the CP
conserving 2HDM.} where $M_{H^+}\gg v$, the mixing angle $\alpha$ (which
diagonalizes the CP-even neutral Mass matrix in the CP-conserving 2HDM) can be
approximated by $\beta-\frac{\pi}{2},$ and the Higgs basis is up to a minus sign
the mass basis,
\be
\bpm h\\H\epm = \bpm -s_{\alpha} & c_{\alpha} \\c_{\alpha} & s_{\alpha}\epm
	\bpm \phi_1 \\ \phi_2 \epm \xrightarrow{\tiny M_{H^+}\gg v}
	\bpm c_{\beta} & s_{\beta}\\ s_{\beta} & - c_{\beta} \epm
	 \bpm\phi_1\\\phi_2\epm
\ee
which according to equation \eqref{eq:higgs_basis} shows that 
\be
h=h_1\qquad H=-h_2.
\ee
Second, since at tree-level in the gauge-eigenstate basis only the 2HDM doublet
$\Phi_2$ couples to top quarks according to our above assumptions, one can
approximate the self-energy corrections to the CP-even Higgs bosons by
\begin{align}\nonumber\label{eq:approx}
	\Sigma^{\mathrm{tops}}_{h_1h_1} = \Sigma^{\mathrm{tops}}_{hh} &= s_\beta^2\Sigma^{\mathrm{tops}}_{\phi_2\phi_2}\\\nonumber
	\Sigma^{\mathrm{tops}}_{h_1h_2} = -\Sigma^{\mathrm{tops}}_{hH} &= c_\beta s_\beta
		\Sigma^{\mathrm{tops}}_{\phi_2\phi_2} = \frac{1}{\tan\beta}\Sigma^{\mathrm{tops}}_{hh} \\
	\Sigma^{\mathrm{tops}}_{h_2h_2} = \Sigma^{\mathrm{tops}}_{HH} &= c_\beta^2 \Sigma^{\mathrm{tops}}_{\phi_2\phi_2} =
		\frac{1}{\tan^2\beta}\Sigma^{\mathrm{tops}}_{hh}.  
\end{align}
Now we consider $v^2(m_t)\lambda^{\mrm{SM}}(m_t)$ to be the leading-log
resummation of the one-loop-leading-log contribution coming from the
$\Sigma^{\mrm{tops}}_{h_1h_1}$ self-energy correction. Therefore, using Eq.~\eqref{eq:approx}, we
incorporate this into the Higgs basis matrix as follows
\be
\mathcal M^{\mathrm{Higgs}}_{\text{approx}} =  \mathcal M^{\mathrm{Higgs}} + \bpm\gamma^{\mrm{resum}} &
\frac{1}{\tan\beta}\gamma^{\mrm{resum}} & 0 & 0\\ \frac{1}{\tan\beta}\gamma^{\mrm{resum}}
&\frac{1}{\tan^2\beta}\gamma^{\mrm{resum}}& 0 &0 \\ 0 & 0 & 0& 0 \\ 0 &0&0 &
0\epm,
\ee
where $\gamma^{\mrm{reum}}\equiv v^2(m_t)\lambda^{\mrm{SM}}(m_t) - v^2(m_{H^+})\lambda^{\mrm{SM}}(M_{H^+}).$ 
Then, the eigenvalues of the matrix 
\be\label{mass_mat}
	\mathcal M^{\mrm{Higgs}}_{\mrm{approx}} - \hat{\Sigma}(p^2,\mu) + \hat{T}
\ee
are calculated, where $\hat{\Sigma}$ is the $4\times4$ matrix of one-loop
self-energies of the neutral Higgs bosons in the gauge basis and $\hat{T}$ is
the matrix of one-loop neutral tadpole corrections in gauge basis given in
appendix \ref{tadpole}, both in the $\overline{\mrm{MS}}$ scheme and using the parameters at $m_t$.

To calculate these eigenvalues, it is necessary to choose the renormalization
scale $\mu$ and the external momentum $p^2$. The renormalization scale is set to
$m_t$, and the external momenta are chosen to be the tree-level masses. We
choose to calculate the eigenvalues one a time, with the mass matrix
diagonalized once for each tree-level mass. For example, the lightest neutral
Higgs corresponds to the lightest eigenvalue of the loop-corrected mass matrix
evaluated with the external momentum set to the tree-level neutral Higgs mass.

In order to proceed along the same lines as in the pure 2HDM case, see \ref{2HDMatMHp}, the matrix $\mathcal M^{\mathrm{Higgs}}_{\text{approx}}$ can be rotated back to the interaction eigenstates using the transformation matrix \eqref{eq:Umatrix}. This results in $\mathcal M^2$ with an additional  contribution to the (2,2) element, \begin{align}\label{eq:Msquaredresum}
\mathcal (M^2_{22})^{\mrm{resum}} =  \mathcal M^2_{22}  + \frac{1}{\sin^2 \beta} \gamma^{\mrm{resum}}. 
\end{align}
Using this new matrix $(\mathcal M^2)^{\mrm{resum}}$ and replacing $\mathcal M^2$ by $(\mathcal M^2)^{\mrm{resum}}$ in Eq.~\eqref{eq:det}, one can calculate the Higgs masses by finding the zeros of the resulting equation up to one-loop order. We find that both approaches lead to nearly the same result.
\end{enumerate}
For options (b) to (d), an evaluation at the scale of the
running top-quark mass $m_t (m_t)$ is performed. In these cases, the running
top-quark mass is calculated iteratively.
\end{enumerate}


\section{Numerical Results}
\label{se:results}
\subsection{Choice of input parameters}\label{sse:inputs}
\paragraph{}
The list of relevant input parameters for the calculation is 
\be
\{\underbrace{y_t,\;y_b,\;g_y,\;g_3,\;g_2,\;v,}_{\text{at }
M_t}\;\underbrace{\tan\beta(M_{H^+}),\;M_{H^+},}_{\text{at }
M_{H^+}}\;\underbrace{A_t,\;A_b,\;\varphi_{M_3},\;\mu}_{\text{at } M_s}\}. 
\ee
As mentioned in Sect.~\ref{se:mass}, these parameters are fixed at different
scales. The soft-breaking parameters of the MSSM and the Higgs mixing parameter
$\mu$ are defined at the scale $M_s$, $\tan\beta$ at the scale $M_{H^+}$, and
the SM input parameters are fixed at the low-scale $M_t$ by current experimental
results. The relevant SM observables needed to define the SM couplings, taken
from \cite{Lee:2015uza,  Buttazzo:2013uya}, are
\be\label{SM_inputs}
\begin{gathered}
\alpha_s(M_z) = 0.1184,\qquad M_t = 173.34\;\mathrm{GeV}, \qquad M_W = 80.384\;\mathrm{GeV},\\ M_Z = 91.1876
	\;\mathrm{GeV},\qquad m_b(m_b) = 4.18\;\mathrm{GeV}, \qquad v^2_{G_F} \equiv \frac{1}{\sqrt{2}G_F} = 246.21971
	\;\mathrm{GeV},
\end{gathered}
\ee
with $G_F$ being the Fermi constant. The values for
$y_b,\;y_t,\;g_y,\;g_2,\;g_3$ are extracted from these observables in
\cite{Lee:2015uza, Buttazzo:2013uya} and given below as running parameters at
the scale $M_t$
\be
\begin{gathered}\label{eq:SMcouplvalues}
 g_3 = 1.1666,\qquad g_2 = 0.64779, \qquad g_y = \sqrt{\frac{3}{5}} g_1 = 0.35830 \\
	y_t = 0.94018,\qquad y_b = 0.0156
\end{gathered}
\ee
where in the conversion the value of the SM Higgs pole mass  $M_h^{\text{SM}} =
125.15$~GeV was used according to Ref.~\cite{Buttazzo:2013uya}.  One must also
determine the running vev $v_{\,\overline{\mathrm{MS}}}$ from the vev $v_{G_F}$,
which is experimentally determined via the Fermi constant, which is measured via
the muon lifetime. We derive the
$\overline{\mathrm{MS}}$ vev from the on-shell vev using 
\be
v^2_{\,\overline{\mathrm{MS}}} = v^2_{\text{OS}} + \delta{v}^2_{\text{OS-finite}}, 
\ee
defining the on-shell vev by
\be
v^2_{\text{OS}} \equiv \frac{4 M^2_W s^2_W}{e^2} = v^2_{G_F}(1+\Delta r) 
\ee
with the counterterm $\delta{v}^2_{\text{OS-finite}}$ given in
Eq.~\eqref{eq:vevOSCT} in the appendix. Here, $s^2_W = 1-\frac{M_W^2}{M_Z^2}$
denotes the sine squared of the weak mixing angle and $\Delta r$ parameterizes
the one-loop radiative corrections to the muon decay in the Fermi Model
\cite{Sirlin:1980prd}, the process by which $v_{G_F}$ is defined. The different
formulas needed for the conversion are collected in the
App.~\ref{se:vevconversion}. The value for $v_{\,\overline{\mathrm{MS}}}$ at the
scale of the top pole mass $M_t$ is then 
\begin{align}\label{eq:SMvevvalue}
	v_{\,\overline{\mathrm{MS}}}(M_t) = 247.3897 \text{ GeV}
\end{align}
employing again the Higgs-boson mass $M_h^{\text{SM}} = 125.15$~GeV. In the
numerical evaluation of the masses of the Higgs bosons and mixings, we use the
numbers given in Eqs.~\eqref{eq:SMcouplvalues}  and \eqref{eq:SMvevvalue} as
input values for the SM parameters\footnote{The conversion from the SM input
values given in Eq.~\eqref{SM_inputs} to the parameters in
Eqs.~\eqref{eq:SMcouplvalues}  and \eqref{eq:SMvevvalue} involves the
Higgs-boson mass so that, since we calculate the mass of the Higgs boson, a more
sophisticated approach would be an iteration where the conversion is
recalculated depending on the obtained result for the Higgs-boson mass. Since in
a physical viable scenario, the SM-like Higgs-boson mass should be about 125
GeV, we consider the ``one time'' conversion as sufficient.}.

For the remaining input parameters $M_{H^+}$, $\tan \beta$, $A_t$, $A_b$, $M_s$,
$\mu$, no measured values exist, but the experimental searches and measurements
constrain the viable parameter space.  The exclusion bounds from searches of
further Higgs bosons  \cite{Sirunyan:2018zut, Aaboud:2017sjh} constrain in
particular the region of high $\tan \beta$  and light ``heavy'' Higgs bosons.
Taking these results together with further studies of different parameter
scenarios \cite{Bahl:2018zmf, Bahl:2019ago}, it is clear that scenarios with
$\tan\beta>10$ and $M_{H^+},M_A <500$ GeV are strongly disfavoured by LHC data.
Flavour observables support these constraints, as discussed in
Ref.~\cite{Arbey:2017gmh} for different types of the 2HDM.  In our numerical
analysis, we therefore favour values for the mass of the charged Higgs boson of
$M_{H^+} \geq 500$ GeV and $\tan\beta = 5$. To show specific features of the
results of our calculation, we will however partly take into account scenarios
that do not fulfill these constraints. 

It should be noted that in the MSSM, some of
the non-vanishing parameter phases can be eliminated by symmetry transformations. Hence, only
certain combinations of phases are physical, i.e.\ can change the value of a
physical observable. Important constraints of these phase combinations come from
electric-dipole moment (EDM) measurements (see
e.g.~Ref.~\cite{Berger:2015eba,Abe:2018qlw,Cesarotti:2018huy} for more recent
studies of the constraints of the MSSM phases due to EDM). Since in our
calculation the phases of the $U(1)$ and $SU(2)$ gaugino-mass parameters $M_1$
and $M_2$ do not enter, we can assume that they are chosen such that the effect
on the EDM is minimized. Furthermore, larger masses of the SUSY particles tend
to relax the constraints coming from the EDM. 

It should be noted that, in this paper, we refrain from explicit checks whether
a certain parameter point is viable, since we focus on specific features of the
results. In particular, using our results at the low scale to study the
constraints from the EDMs for the MSSM phases at the high scale will be
interesting but is left for future work. 

Our default scenario is
\begin{align}\nonumber
A_t &= A_b \equiv A; \quad |A| = |\mu| = 3M_s; \quad M_{H^+} = 500 \text{ GeV}; \quad \tan \beta = 5; \\
\varphi_{A_t} &= \varphi_{A_b} = 2.1 \approx 0.67\pi = 120^\circ; \quad  \varphi_\mu = 0; \quad \varphi_{M_3} = 0.
\end{align}
where $M_s$ is varied.  The choice $|A| = |\mu| = 3M_s$ leads to large threshold
corrections to the 2HDM couplings and maximizes the amount of CP-violation
introduced into the theory. This way we can give an estimate of the largest
effects that can occur. Similarly,  we observe that $\varphi_{A_t} =
\varphi_{A_b} =  2.1 \approx 0.67 \pi = 120^\circ$ maximizes roughly the size of the CP-odd
component of the lightest neutral Higgs boson, see
Sect.~\ref{sse:CPodd_component}. We will however deviate from this default
scenario in order to study the different characteristics and state that
explicitly.

\subsection{Influence of the Running of Complex Parameters}
\paragraph{} 
In this work, we exploit the two-loop RGEs for the 2HDM with each Higgs doublet
coupling to up- as well as down-type fermions including all phases, see
Sect.~\ref{se:mass} for the details of our calculation of the RGEs. The first
numerical results exemplify the effect of including this phase dependence versus
the ``real RGE'' approximation where the phase dependence is taking into account
only via the threshold effects and the phases are assumed to be unaffected by
the running.

\begin{figure}
	\centering
	\begin{subfigure}[]{0.3\textwidth}
		\includegraphics[width=\textwidth]{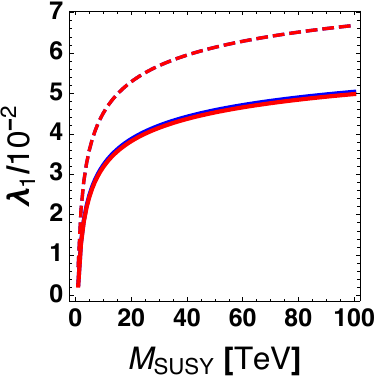}
		\caption{$\lambda_1$}
		\label{complex_real_lam1}
	\end{subfigure}
	\begin{subfigure}[]{0.3\textwidth}
		\includegraphics[width=\textwidth]{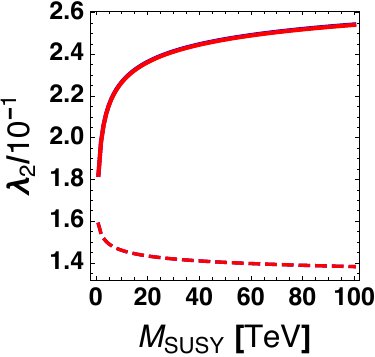}
		\caption{$\lambda_2$}
		\label{complex_real_lam2}
	\end{subfigure}
	\begin{subfigure}[]{0.3\textwidth}
		\includegraphics[width=\textwidth]{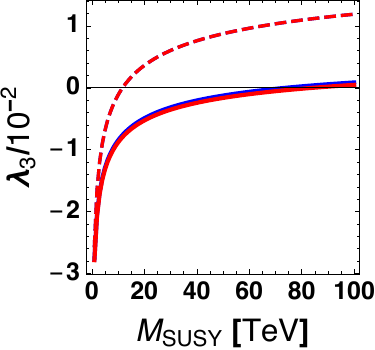}
		\caption{$\lambda_3$}
		\label{complex_real_lam3}
	\end{subfigure}
	\begin{subfigure}[]{0.3\textwidth}
		\includegraphics[width=\textwidth]{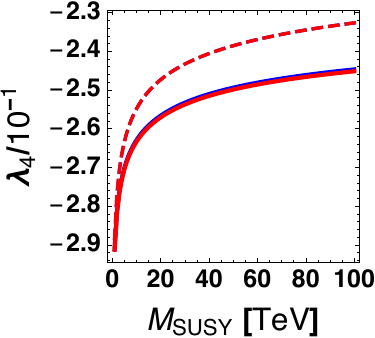}
		\caption{$\lambda_4$}
		\label{complex_real_lam4}
	\end{subfigure}
	\begin{subfigure}[]{0.3\textwidth}
		\includegraphics[width=\textwidth]{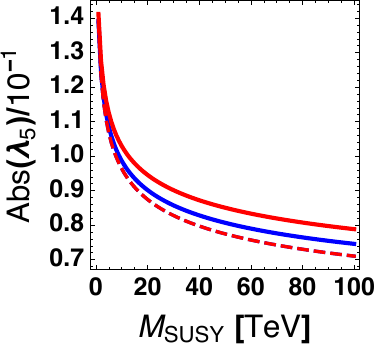}
		\caption{$|\lambda_5|$}
		\label{complex_real_lam5real}
	\end{subfigure}
	\begin{subfigure}[]{0.3\textwidth}
		\includegraphics[width=\textwidth]{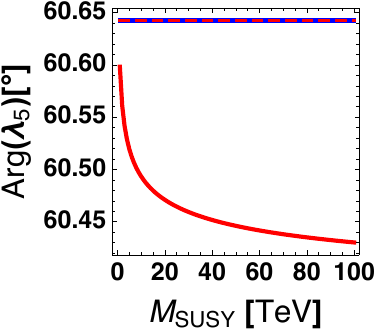}
		\caption{$\arg(\lambda_5)$}
		\label{complex_real_lam5im}
	\end{subfigure}
	\begin{subfigure}[]{0.3\textwidth}
		\includegraphics[width=\textwidth]{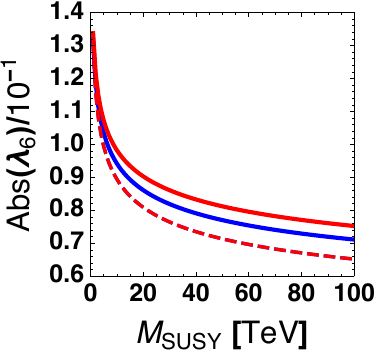}
		\caption{$|\lambda_6|$}
		\label{complex_real_lam6real}
	\end{subfigure}
	\begin{subfigure}[]{0.3\textwidth}
		\includegraphics[width=\textwidth]{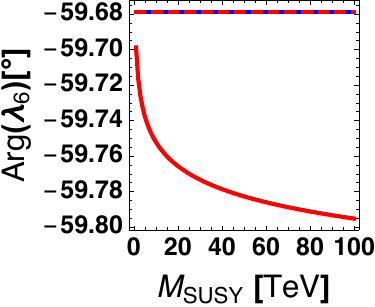}
		\caption{$\arg(\lambda_6)$}
		\label{complex_real_lam6im}
	\end{subfigure}
	\begin{subfigure}[]{0.3\textwidth}
		\includegraphics[width=\textwidth]{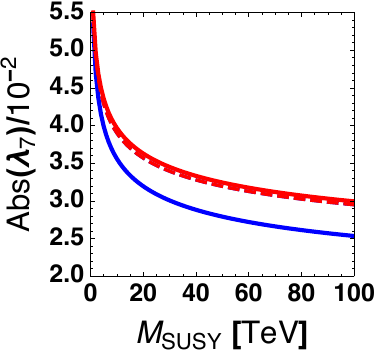}
		\caption{$|\lambda_7|$}
		\label{complex_real_lam7real}
	\end{subfigure}
	\begin{subfigure}[]{0.3\textwidth}
		\includegraphics[width=\textwidth]{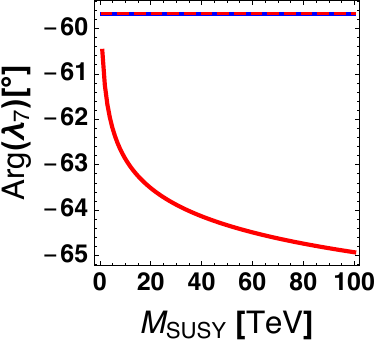}
		\caption{$\arg(\lambda_7)$}
		\label{complex_real_lam7im}
	\end{subfigure}
	\caption{The quartic couplings'
	dependence on $M_s$ for the default scenario: $\tan\beta=5$,
	$\varphi_{A}=2.1 \approx 120^\circ$, $\varphi_{\mu} = \varphi_{M_3}=0$,
	$|\mu|=|A|=3M_s$, $M_{H^+} = 500$ GeV. The red curves are the result of
	employing complex RGEs, and the blue real RGEs.  Dashed lines
  are the couplings at the high scale $M_s$, and solid lines are couplings at
  the scale $M_{H^+}.$
	}
	\label{complex_real_rges}
\end{figure}

In Fig.~\ref{complex_real_rges}, we show the effect of including these phases on
the running of the 2HDM quartic couplings. The values of the couplings are
plotted against the scale $M_s$, demonstrating how this dependence changes when
the running of both the real and the imaginary part of the couplings is taken
into account. The coupling values shown are the ones that either enter the final
evolution at the scale $M_s$ or are calculated via the final evolution to the
scale of the charged Higgs mass in step \ref{masscalculation}. The red lines
represent the values obtained with RGEs taking phases into account while the
blue ones represent values obtained using only real parameters, determining the
sign via the argument of the corresponding parameter at the scale $M_s$. The
dashed lines are the parameters at the scale $M_s$, while the solid lines are
those at $M_{H^+}$. One can clearly see a dependence on whether the running of
the phases is taken into account or not. It changes the resulting absolute value
of the couplings $\lambda_5$, $\lambda_6$ and $\lambda_7$ as well as the phases
themselves. The dependence of the running on the phases is relatively small;
only the phase of $\lambda_7$ shows a change of up to a couple of degrees.
Hence, the overall dependence of the phases determined using the RGEs for the
complex case on the value of $M_s$ is also relatively small. Comparing the
phases at $M_s$ (dashed lines) with the ones obtained with using the real RGEs
(blue), one can double-check that indeed the phases do not change when
exploiting the real RGEs. The absolute values of $\lambda_5$ and $\lambda_6$
change more when the complex RGEs are applied compared to a result with only
real RGEs. The opposite is true for the absolute value for $\lambda_7$, which
changes less if the complex RGEs are applied.

Since the phase values at the low scale enter in the prediction of the
EDM, they  can be relevant for checking the exclusion of high-scale CP-violating
MSSM scenarios due to the measurements of the EDM. 

\subsection{Comparison of Methods for Computing Masses}
\label{sse:comparison}
\paragraph{}
In Sect.~\ref{mass_calculation}, the calculation procedure was explained, and
in step \ref{masscalculation} of this procedure we discussed several possibilities for
calculating the mass of the Higgs boson $m_h$ at the low-energy scale: a)
 exploiting the 2HDM at the scale $M_{H^+}$, b) employing
the 2HDM at the scale $m_{t}$, c) matching the 2HDM to the SM and using the SM
at the scale $m_t$, or d) approximating the effects of the matching to the SM
and running down to scale $m_t$.  Using the first two approaches,  one keeps
information about all the Higgs masses and their mixings with one another. This
is very important, as we are also interested in the size of the CP-odd component
of the lightest Higgs boson. However, if the scale $M_{H^+} \gg v \sim m_t$, we
again encounter large logarithms. The masses of the heavy Higgs bosons will not
pose a problem, since the logarithms are not equally enhanced by large prefactors as the ones appearing in the calculation of light Higgs mass and since the relative shifts due to the large tree-level masses are smaller, so we can trust the perturbative results for
these masses without an additional resummation of logarithms. These large
logarithms are more important for the lightest Higgs, on the other hand.
\begin{figure}
	\centering
	\begin{subfigure}[]{.4\textwidth}
		\includegraphics[width=\textwidth]{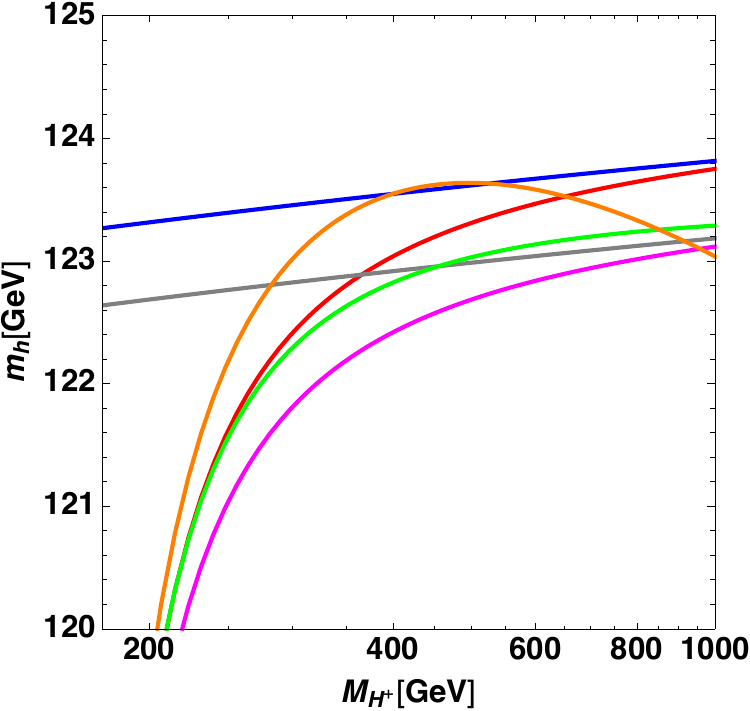}
		\caption{$\tan\beta=5$}
	\end{subfigure}
	\begin{subfigure}[]{.4\textwidth}
		\includegraphics[width=\textwidth]{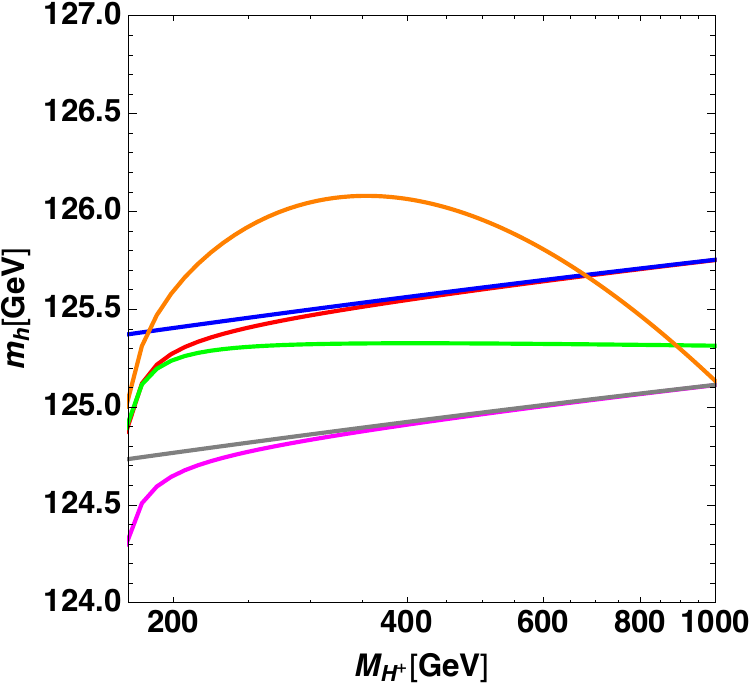}
		\caption{$\tan\beta=10$}
	\end{subfigure}\vspace{5mm}
	\begin{subfigure}[]{.7\textwidth}
		\includegraphics[width=\textwidth]{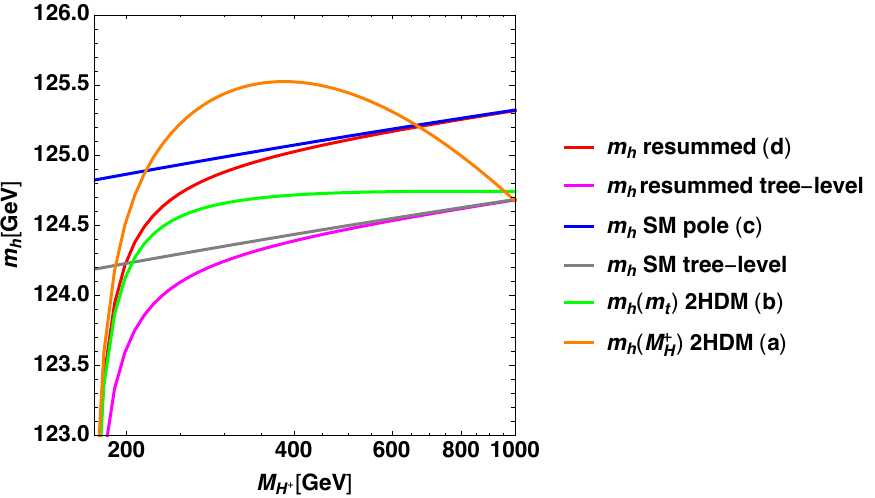}
		\caption{$\tan\beta=20$}
	\end{subfigure}
	\caption{The mass of the light Higgs boson is shown in dependence on the
	scale $M_{H^+}$ for the scenario $|A|=|\mu|=3M_s$, $M_s = 3$ TeV, $\varphi_A
	= \varphi_\mu = \varphi_{M_3} = 0$, $\tan \beta = 5, 10, 20$ employing
	different approximations. The curves labeled ``resummed" present the result
	where the logarithms of the form $\ln(M_{H^+}/M_t)$ have been resummed by
	matching to SM, but the Higgs mass is still calculated using the mass matrix
	of the 2HDM. Those labeled with ``SM" are calculated by decoupling
	completely from the 2HDM, and calculating a purely SM mass. Those labeled
	with ``2HDM" refer to those calculated treating the 2HDM as the low-energy
	theory.  For the ``tree-level" curves, one-loop corrections have not been
	included in the calculation of the pole mass.}
	\label{method_compare}
\end{figure}

Figure~\ref{method_compare} shows the result of calculating the lightest
Higgs-boson mass using the different approaches for real parameters. The
calculation of the mass of the lightest Higgs boson using the 2HDM at the scale
$M_{H^+}$ and $m_t$ is denoted with ``$m_h(M_{H^+})$ 2HDM (a)'' and ``$m_h(m_t)$
2HDM (b)'', respectively. The result where the SM is the low-energy theory is
called ``$m_h$ SM pole (c)''. A variant of this result is ``$m_h$ SM pole tree
level'' where the self-energy contribution  $\hat{\Sigma}^\text{SM}$  as well as
the one-loop tadpole contribution $\hat{T}^{\text{SM}}$ in Eq.~\eqref{eq:polemass}
are neglected. Finally, the results labeled ``$m_h$ resummed (d)'' correspond to
the approximation where parts of the logarithms $\ln\left(M_{H^+}/m_t\right)$
are resummed. The result ``$m_h$ resummed tree level'' neglects the self-energy
and the tadpole contributions in Eq.~\eqref{mass_mat}. 

For large $M_{H^+}$, the ``$m_h$ SM pole" (c) is expected to be the most
``correct" answer for large $M_{H^+}$, while the result obtained by using the
2HDM as the EFT should be the best result when $M_{H^+}\sim v \sim m_t$.
The two 2HDM results agree for low $M_{H^+}$ but start to deviate quickly
with rising $M_{H^+}$. In the self-energy contribution in Eq.~\eqref{eq:det},
logarithms of $\ln\left(M_{H^+}/m_t\right)$  arise if the self energy is
evaluated at the scale $M_{H^+}$. Enhanced via large top Yukawa couplings, this
result differs quickly from the other 2HDM result where these logarithms vanish
in the self energy due to the scale choice and is taken into account via the
running of the parameters. Therefore, it is preferable to calculate the Higgs
mass $m_h$ at the scale $m_t$. If $M_{H^+}$ however is large, $M_{H^+} \gg m_t$,
the heavy Higgs-bosons have to be decoupled from the running of the parameters
and the SM is the correct low-energy theory. For $M_{H^+} = 1000$ GeV, the
deviation of the ``$m_h$ SM pole (c)" result from the ``$m_h(m_t)$ 2HDM (b)''
result is roughly 500 MeV for all values of $\tan \beta$.
 That means that, for $M_{H^+} <
1000$ GeV, the 2HDM  can still be used as a reasonable low-energy theory but
with increasing theoretical uncertainty for increasing values of $M_{H^+}$. The difference between both results ``$m_h$ SM pole (c)" and ``$m_h(m_t)$ 2HDM (b)'' first decreases and then starts growing. For $\tan \beta = 10$ and $\tan \beta = 20$, the increase sets already in for lower values of $M_{H^+}$. 

  For
small $M_{H^+}$,  the mixing of the Higgs bosons becomes relevant, which
leads to a decrease of the Higgs-boson mass in the 2HDM results. The ``$m_h$ SM
pole (c)" result does not take the mixing of the Higgs bosons into account due
to the decoupling of the heavy Higgs bosons. The ``$m_h$ resummed (d)'' result
follows nicely the 2HDM result for low values  and the SM result for large
values of $M_{H^+}$. Hence, it interpolates well between the two options.
Therefore, we have chosen this as default option. 

In addition, the effect of the one-loop self energies in the calculation of the
pole masses can be read off when comparing the ``$m_h$ resummed (d)'' and
``$m_h$ resummed tree level'' result (or similar the corresponding SM pole mass
results). The top and bottom quark contributions that we take into account lead
to approximately a 0.6 GeV rise of the result.
 
\subsection{The Mass of the Light Higgs Boson}
\paragraph{}
In this section, we discuss the dependence of the lightest Higgs boson on the
MSSM input parameters $\mu$, $A_t=A_b=A$, $\tan\beta,$ and $\varphi_{M_3}$.

\begin{figure}[t]
	\centering
	\begin{subfigure}[]{.4\textwidth}
		\includegraphics[width=\textwidth]{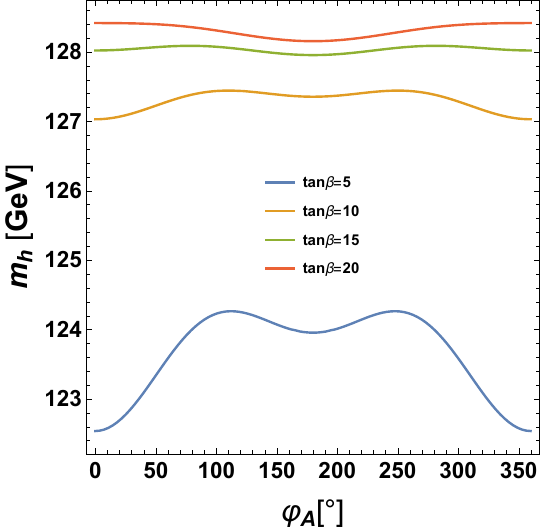}
		\caption{$|A| = |\mu| = 2M_s$}
	\end{subfigure}
	\begin{subfigure}[]{.4\textwidth}
		\includegraphics[width=\textwidth]{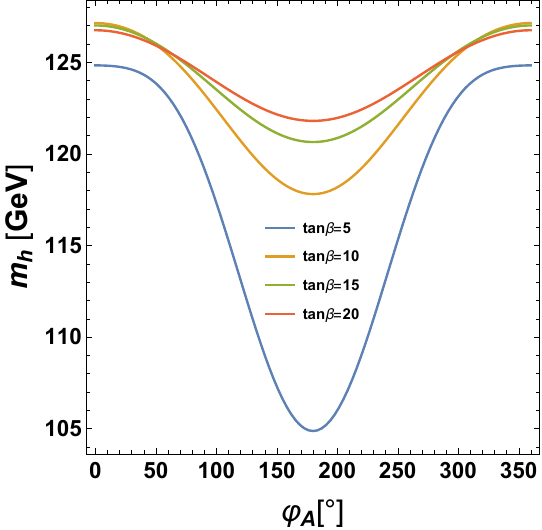}
		\caption{$|A| = |\mu| = 3M_s$}
	\end{subfigure}\vspace*{0.3cm}
	\begin{subfigure}[]{.4\textwidth}
		\includegraphics[width=\textwidth]{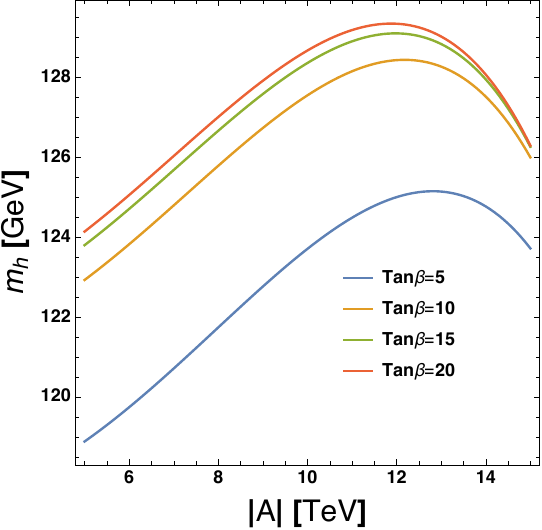}
		\caption{$\varphi_A = 0$, $|\mu|=M_s$}
	\end{subfigure}
	\begin{subfigure}[]{.4\textwidth}
		\includegraphics[width=\textwidth]{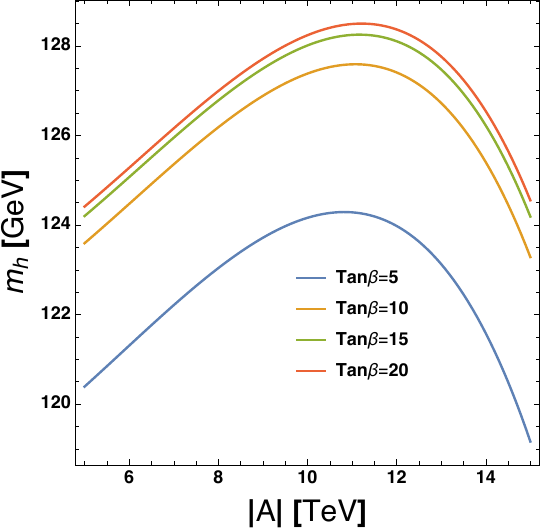}
		\caption{$\varphi_A = 2.1 \approx 120^\circ$, $|\mu|=M_s$}
	\end{subfigure}
	\caption{The upper row presents the mass of the lightest Higgs boson depending on the phase
	$\varphi_A$ while the lower role shows the dependence on the absolute value
	of $A$ for a scenario with $M_{H^+}=500$ GeV, $M_s = 5$ TeV,
	$\varphi_{M_3}=0$, and $\tan \beta = 5, 10, 15, 20$.}
	\label{fig:A_dependence}
\end{figure}

The first row of Fig.~\ref{fig:A_dependence} shows the dependence on the common
phase $\varphi_A$. Firstly, the plots demonstrate that the sensitivity of
the mass of the lightest Higgs boson to the phase is highly dependent on
$\tan\beta$, as the mass fluctuates more with $\varphi_A$ for low values of
$\tan\beta$. For $\tan \beta = 5$, the mass of the Higgs boson leads to a change
of the Higgs-boson mass of almost 20 GeV in the quite extreme case of
$|A|/M_s =3$, while for $\tan \beta = 20$ it is only about 5 GeV for otherwise
the same parameters. Secondly, it can be seen that the quantitative features are
strongly dependent on the ratio
$r_{\mathrm{inputs}}\equiv\frac{A_b,A_t,\mu}{M_s}$. The qualitative dependence
on the phase is in fact opposite for the two cases $r_{\mathrm{inputs}} =2$ and
$r_{\mathrm{inputs}} =3$, where the mass has either a minimum for
$r_{\mathrm{inputs}} =3$ or near maximum value for $\varphi_A = 180^\circ$ for
$r_{\mathrm{inputs}} = 2$. The different dependence on the phase $\varphi_A$
with respect to the ratio $|A|/M_s$ can also be read off the second row of
Fig.~\ref{fig:A_dependence} where the lightest Higgs mass is plotted against the
magnitude of the common trilinear $|A|$ with $\mu$ fixed to $\mu = M_s$. Here,
it is seen that the mass peaks for $|A| \approx 12.5 \text{ TeV } = 2.5 M_s $ for $\varphi_A
= 0$ (left plot), and increasing $\tan\beta$ shifts the peak slightly to lower
values of $|A|/M_s$. For $\varphi_A = 2.1 \approx 120^\circ$, the peak is shifted to a lower
value $|A|/M_s\approx 2.2$ while increasing $\tan \beta$ leads to a shift to
slightly higher values in this case. Comparing the left and right plot of the lower row
of Fig. ~\ref{fig:A_dependence}, one can read off for which $|A|/M_s$, the
values of the Higgs-boson mass are smaller for $\varphi_A = 0$ than for
$\varphi_A = 180^\circ$ and vice versa resulting in the changed maxima in the
upper row of Fig.~\ref{fig:A_dependence}. 

The effect of changing $|\mu|$ can be seen by comparing the upper rows from
Fig~\ref{fig:A_dependence} with the lower rows. The value of $\mu$ changes from
$\mu = 2 M_s$ ($\mu = 3 M_s$) to $\mu=M_s$ from the top left (right) figure to the bottom row. Comparing
values for $m_h$ for  $\varphi_A = 0^\circ$ in the upper left (right) plot with the ones for $|A| = 10$ TeV ($|A| = 15$ TeV) in the lower left plot leads to a rise of the Higgs-boson mass by roughly 1 GeV. For $\varphi_A = 2.1 \approx
120^\circ$, the same change leads to a shift of up to 6 GeV in the example
scenarios, which is reached for $\tan \beta = 5$.

\begin{figure}
	\centering
	\includegraphics[width=.5\textwidth]{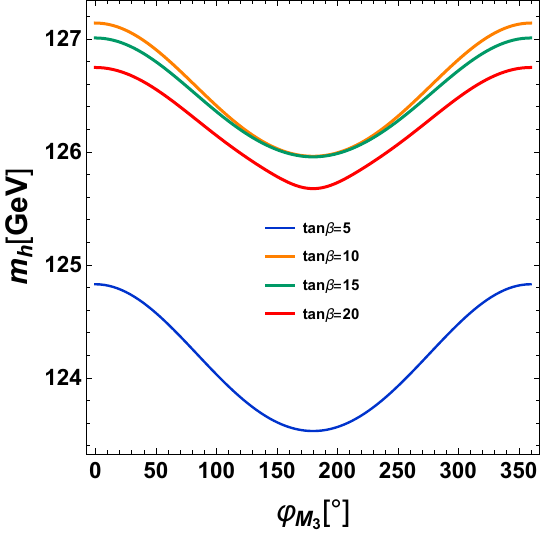}
	\caption{Dependence of the mass of the lightest Higgs boson on the phase
	$\varphi_{M_3}$  for a scenario with $M_{H^+}=500$ GeV, $\varphi_{A} = 0$, $M_s = 5$ TeV, and $|A|=|\mu|=3M_s$.}
	\label{Phi_M3}
\end{figure}

Figure \ref{Phi_M3} shows the dependence on $\varphi_{M_3}$. The overall
dependence on $\varphi_{M_3}$ is not very large and varying $\varphi_{M_3}$
leads to changes of the order of 1 GeV. The largest shift of 1.3 GeV in the
considered scenarios can be found for $\tan \beta = 5$ and similar for $\tan
\beta = 20$. For $\tan\beta = 10, 15$, the changes are slightly smaller.

\newpage
\subsection{A CP-odd Admixture to the Light Higgs boson}\label{sse:CPodd_component}
\paragraph{}
\begin{figure}
	\centering
	\begin{subfigure}[t]{.4\textwidth}
    \includegraphics[width=\textwidth]{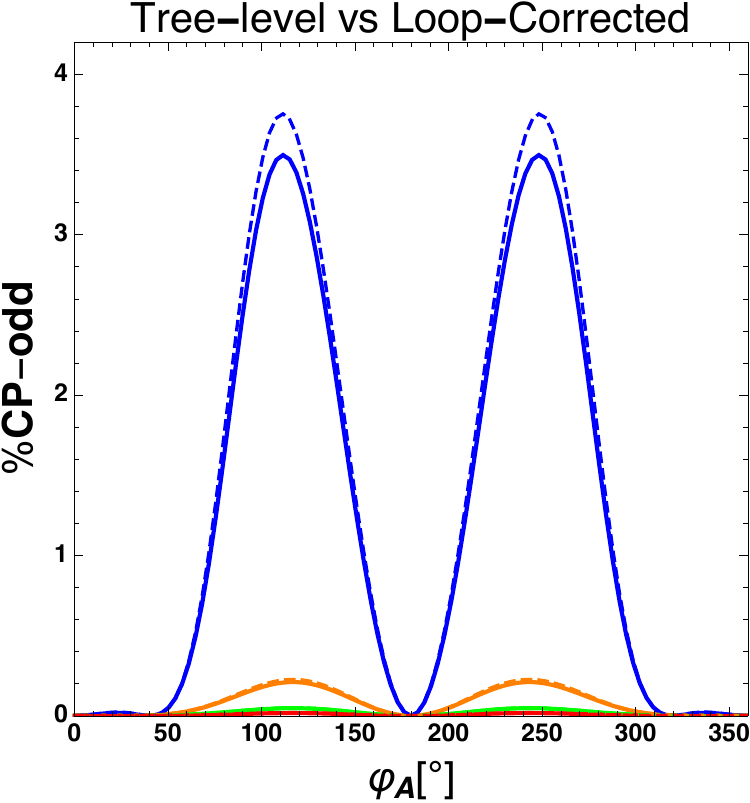}
    \caption{The solid curves are calculated using the tree-level mixing matrix,
and the dashed curves are calculated using the one-loop mixing matrix at
zero-external momentum ($p^2 = 0$), both calculated at $\mu=M_{H^+}$}
   \end{subfigure}
   \hspace{.5cm}
	\begin{subfigure}[t]{.4\textwidth}
    \includegraphics[width=\textwidth]{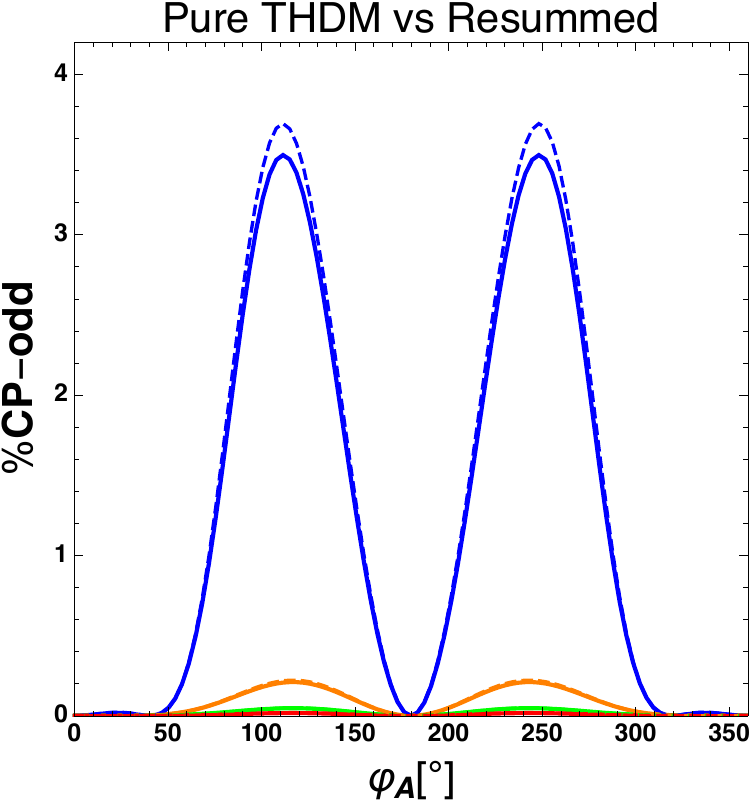}
    \caption{The solid curves are calculated using the tree-level mass matrix
without the resummed log contribution, and the dashed curves include the log
resummation, both at $\mu = M_{H^+}$}
   \end{subfigure}
    \caption{The other parameters of the scenario are $\tan\beta=5$, $M_s=30$ TeV, $\varphi_\mu = \varphi_{M_3} = 0$, and $|A| = |\mu| = 3M_s.$ } 
    \label{CPodd_mh_phiA}
\end{figure}

\begin{figure}
	\centering
	\includegraphics[width=.7\textwidth]{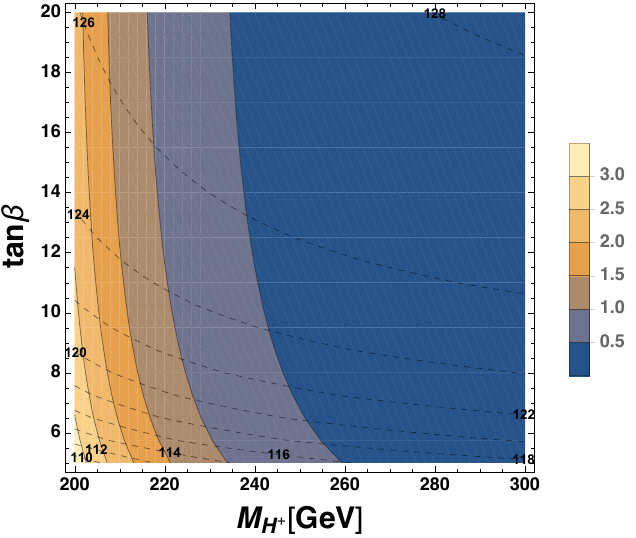}
	\caption{The dependence of the size of the CP-odd component on $M_{H^+}$ and $\tan \beta$ is shown for a scenario $M_s=30$ TeV,$\varphi_{M_3} = 0,$ $\varphi_{A} = 2.1 \approx 120^\circ$, and $|A| = |\mu| = 3M_s$. The
    color coding gives the percentage of the lightest Higgs boson that is CP-odd.}
	\label{CPodd_tanb_MHplus}
\end{figure}

The detection of a CP-odd component in the SM-like Higgs boson would be a sign
of new physics, and it is therefore interesting to explore the size of such a
component generated by the CP-violating phases in the complex MSSM. While we
leave a detailed analysis of the current experimental sensitivity to this CP-odd
component for future work, in this section we aim to give a qualitative picture
of the size of this component and its dependence on the relevant parameters.

We calculate the CP-odd component using the tree-level mixing matrix
$\mathcal{M}^2$ of ~\eqref{eq:2HDMmassmatrixneutral}, inserting the parameter
values at $M_{H^+}$ obtained after the final iteration. First, we separate out
the Goldstone field by rotating by $D\mathcal{M}^2D^{\mrm{T}}$, where
\be
D= \bpm I_2 & 0 \\ 0 & \begin{matrix} -\sin\beta &
  \cos\beta \\  \cos\beta & \sin\beta \end{matrix}\epm 
\ee
with $I_2$ being the two-dimensional identity matrix.
This rotates the mass matrix into the 
basis of the fields $\phi_1, \phi_2, a, G$, where $\phi_1,\phi_2$ are pure CP-even
fields, $a\equiv-a_1\sin\beta
+ a_2\cos\beta$ is pure CP-odd, and $G$ is the the Goldstone boson. In this
basis, the mass matrix is block-diagonal, with a $3\times3$ block for the fields
$\phi_1,\phi_2,a$, and a $1\times1$ block for the Goldstone boson that does not mix with any of
the other fields. 
 The rotation matrix $P$ that diagonalizes the $3\times3$ matrix relates the
physical fields, $h$, $H_2$, $H_3$ to the interaction fields after electroweak symmetry breaking,  $\phi_1$,  $\phi_2$,  $a$,  by
\be \label{eq:mixing} 
\bpm h \\ H_2 \\ H_3 \epm = P\bpm \phi_1 \\ \phi_2 \\ a \epm.
\ee
The third column of $P$ gives the size of the CP-odd component of each of the
physical fields. We choose to plot the CP-odd percentage rather than the
mixing-matrix component itself, which is just $P_{i3}^2*100$ for $i, 1...3$. 

In Fig.~\ref{CPodd_mh_phiA}, we show how the CP-odd component depends on
$\varphi_A$ and $M_{H^+}$. The solid lines in Fig.~\ref{CPodd_mh_phiA} (a) and
(b) are the same and depict the result of the procedure just described. It is
seen that this component is maximized for values of the phases from $110^\circ$ to $120^\circ$,
justifying our claim from Sect.~\ref{sse:inputs}.  In addition, the results
confirm that the CP-odd component very quickly drops to vanishingly small values
as $M_{H^+}$ increases. This agrees with previous findings, see e.g. Ref.~\cite{Li:2015yla}. Since data
disfavours a charged Higgs boson with large mass, it also disfavours a sizeable
CP-odd component of lightest Higgs boson in the MSSM. 

Figure~\ref{CPodd_mh_phiA}  (a) also shows the difference between the one-loop
corrected mixing matrix evaluated at zero external momentum ($p^2=0$) using the
mass matrix given in Eq.~\eqref{mass_mat} and the tree-level case, both with parameters at the scale $M_{H^+}$, which is
evidently numerically small. The largest effect can be seen for $M_{H^+} =
200$~GeV in the peak region where the  inclusion of one-loop corrections leads
to an increase of the squared CP-odd component of about 0.2 percentage points.
If not otherwise stated, we use the tree-level mixing in the following.  

In Fig.~\ref{CPodd_mh_phiA} (b), we compare the approach described above (solid
line) with the approach where the mass matrix
$(\mathcal M^2)^{\text{resum}}$ from Eq.~\eqref{eq:Msquaredresum} is used to
calculated the CP-odd admixture (dashed line). The maximal difference is again
about 0.2 percentage points and decreases with larger values of $M_{H^+}$. Not
shown are one-loop effects in the resummed approach as well as results when
parameters are evaluated at the scale $m_t$. The corresponding results would lie
approximately between the dashed lines in Fig.~\ref{CPodd_mh_phiA} (a) and (b).

Figure \ref{CPodd_tanb_MHplus} shows the percentage of the lightest Higgs boson
that is CP-odd (shaded contours) and the mass of lightest Higgs boson (dashed
line contours) in the $(M_{H^+},\tan\beta)$ plane. As before, the CP-odd
component is calculated according to the approach described above using
$\mathcal M^2$ and parameters evaluated at the scale $M_{H^+}$ without including
loop effects. Again, the CP-odd component drops rapidly with increasing
$M_{H^+}$, and one sees that it is only weakly dependent on $\tan\beta$. In
light of the discussion in Sect.~\ref{sse:inputs} and the mass of the observed
Higgs boson being $\sim~125$ GeV, the results of this section suggest that a
large CP-odd component of the lightest Higgs boson is strongly disfavoured by
experimental data. Already for $M_{H^+} = 260$~GeV in this scenario the CP-odd
component drops below 0.5\%.

According to
Ref.~\cite{Berge:2015nua}, an angle $\varphi_\tau$ of about $4^\circ$ can be
reached with the high-luminosity run of the  LHC. The angle $\varphi_\tau$ is
defined via the effective Lagrangian given in four-component Dirac notation
\begin{align}
  \mathcal{L}^{\tau-\text{Yukawa}}_{\text{eff}} = - \frac{m_\tau}{v}
\kappa_\tau\left(\cos\varphi_\tau \bar{\tau} \tau +\sin\varphi_\tau \bar{\tau}
\text{i} \gamma_5 \tau\right) h 
\end{align} 
where $\kappa_\tau$ denotes the
change of the absolute strength of Yukawa coupling with respect to the SM while
$\varphi_\tau$ governs the amount of CP-violation. To connect to the notation of
\eqref{yukawas}, we write this in two-component notation as
\begin{equation}
  \label{eq:effective_tau_yukawa}
  \mathcal{L}^{\tau-\text{Yukawa}}_{\text{eff}} =
-\frac{m_\tau}{v}\kappa_\tau\left(\cos\varphi_\tau \tau\tau_c -
\text{i}\sin\varphi_\tau \tau\tau_c \right)h + \mrm{h.c.}
\end{equation}
In the case of the 2HDM discussed here, assuming that the coupling of the second
Higgs doublet to the tau leptons is negligible, $\tan \varphi_\tau =
-s_\beta P_{13}/P_{11}\approx - P_{13}/P_{11}$ for scenarios with
$t_\beta\gtrapprox 5$. A CP-odd component of $P_{13}^2\star100 = 0.5\%$ as discussed
above will lead to $\varphi_\tau\gtrapprox4^\circ$, since $P_{11}
\leq 1$. Going back to the scenarios discussed in Fig.~\ref{CPodd_mh_phiA}, the
maximal value of $\varphi_\tau$ is larger than $20^\circ$ for $M_{H^+} =
200$~GeV and larger than $3^\circ$ but below $4^\circ$ for $M_{H^+} = 500$~GeV.
Hence, these values suggest that if the MSSM is the final answer, then while it
will certainly be difficult, it might be possible to determine a CP-odd component experimentally with further
improvements of the measurements. However, we neither checked whether the Higgs signal rates are in the
experimentally allowed regime---this is particularly relevant if $\varphi_\tau$
is enhanced by a smaller $P_{11}$ as it is actually the case for the scenario in
Fig.~\ref{CPodd_mh_phiA} for $M_{H^+} = 500$~GeV---nor took into account any
constraints from the electric dipole moments. Thus, to find out whether there is
still a viable region of parameter space with a measurable CP-odd admixture of
the SM-like Higgs boson in the MSSM with heavy superpartners, further
investigations are required.

\newpage
\subsection{Masses and Mixings of the Heavy Higgs Bosons}
\paragraph{}
\begin{figure}
	\centering
	\includegraphics[width=.5\textwidth]{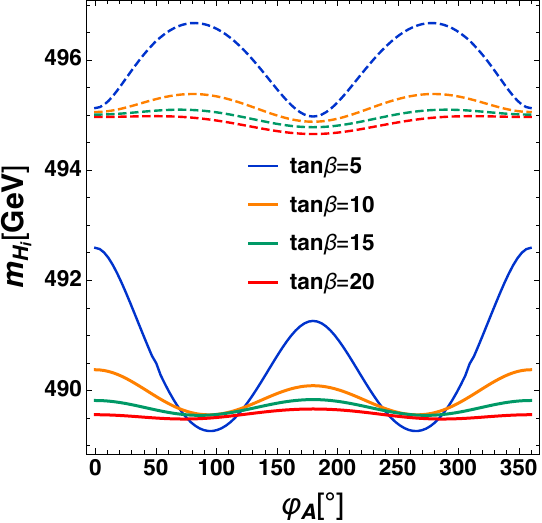}
	\caption{The dependence of the mass of the second lightest  and the heaviest Higgs boson on the phase $\varphi_A$ is shown for the parameter scenario $M_s= 30$ TeV, $M_{H^+}=500$ GeV, $|A|=|\mu| = 3M_s$, $\varphi_{M_3}
= \varphi_\mu = 0$. The solid lines correspond to the mass of the second lightest Higgs boson, $m_{H_2}$, and the dashed lines
to the mass of the heaviest Higgs boson, $m_{H_3}$.}
	\label{heavyhiggs_vs_argA}
\end{figure}

\begin{figure}
	\centering
	\begin{subfigure}[]{.4\textwidth}
		\includegraphics[width=\textwidth]{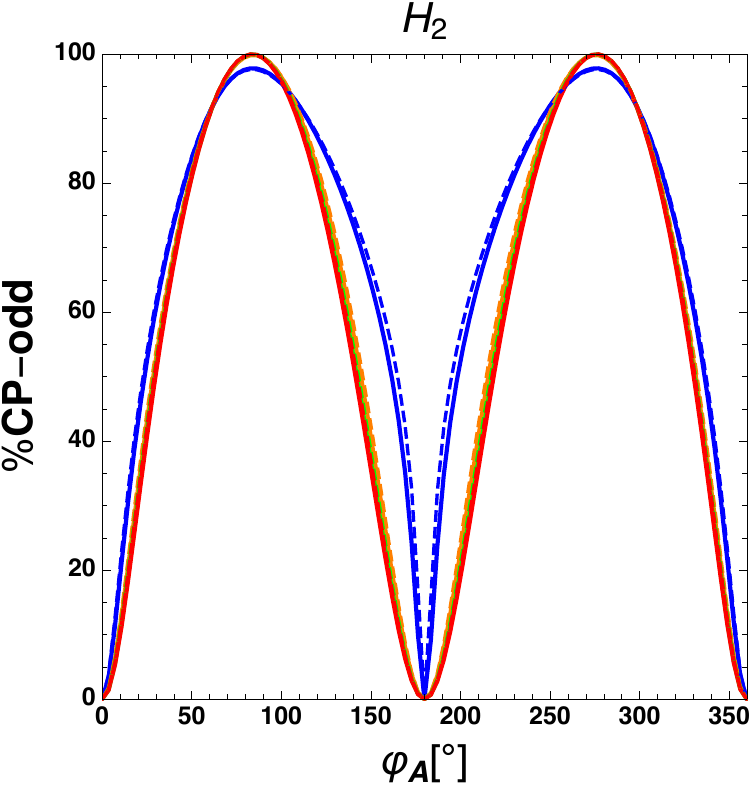}
	\end{subfigure}
	\begin{subfigure}[]{.4\textwidth}
		\includegraphics[width=\textwidth]{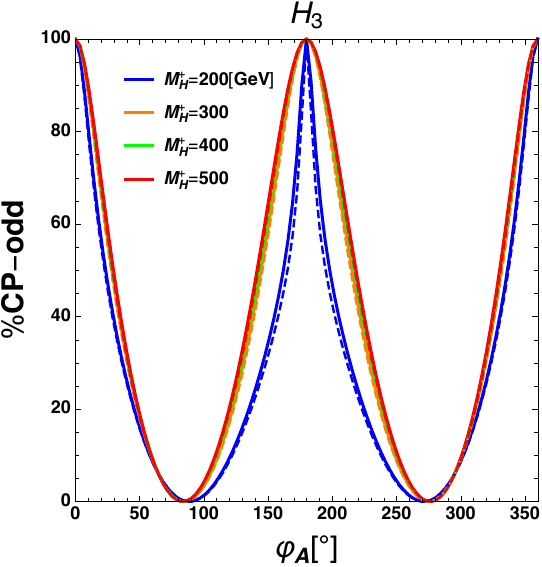}
	\end{subfigure}
	\caption{The dependence of the CP-odd component in percent for the next-to lightest Higgs boson, $P_{23}^2\star 100$, in the left plot and of the heaviest Higgs boson, $P_{33}^2\star100$, in the right plot on the phase $\varphi_A$ for the scenario $\tan\beta=5$, $M_s=30$ TeV, $\varphi_\mu = \varphi_{M_3} = 0$, and
	$|A| = |\mu| = 3M_s.$ The solid curves are calculated using the tree-level
	mixing matrix, and the dashed curves are calculated using the one-loop
	mixing matrix at zero-external momentum ($p^2 = 0$).}
	\label{mixing_phiA}
\end{figure}
In this section, we explore the masses and mixings of the heavy Higgs bosons
$H_2$ and $H_3$. The masses are calculated according to
Sect.~\ref{mass_calculation} where, in step~\ref{masscalculation} of the calculation procedure,
option (d) is exploited where large logarithms $\ln(M_{H^+}/m_t)$ are resummed.  In Fig.~\ref{heavyhiggs_vs_argA}, the dependence of the masses of the two
heavy neutral Higgs bosons $m_{H_2}$ and $m_{H_3}$ on
$\varphi_A$ is shown with solid and dashed lines, respectively.  Similar to the mass
of the light Higgs boson, the phase dependence is larger for low $\tan \beta$,
with a change of the mass of the second-to lightest neutral Higgs boson of
roughly 3.5 GeV and of the heaviest neutral Higgs boson of 1.5 GeV for $\tan \beta
= 5$. This phase dependence is gradually washed out for increasing $\tan\beta$. 

 In Fig.~\ref{mixing_phiA}, the CP-odd percentage is shown with
 $P_{23}^2$ and $P_{33}^2$ defined in Eq.~\eqref{eq:mixing} and evaluated at
 tree level with the 2HDM parameters at the scale $M_{H^+}$. Figure~\ref{mixing_phiA}
 demonstrates that the mixing of the two heavy neutral Higgs bosons is very
 nearly independent of $M_{H^+}$. Only for $M_{H^+}= 200$ GeV can slight deviations
 from the other results be observed. For $M_{H^+}= 200$ GeV, the CP-oddness
 is also shared with the lightest neutral Higgs boson, while for larger $M_{H^+}$
 the CP-odd component is mostly shared between the second lightest and heavy
 Higgs boson. One can also read off of Fig.~\ref{mixing_phiA} that for real
 values of $|A|$ i.e. for $\varphi_A = 0,180, 360^\circ$, the heavy Higgs boson
 $H_3$ is CP-odd and $H_2$ is CP-even.  However, for $\varphi_A
 \approx 90, 270^\circ$, the next-to lightest neutral Higgs boson is mostly
 CP-odd. That means both Higgs bosons oscillate between $\sim 100\%$ CP-odd
 and CP-even with the phase $\varphi_A$. 
 
 In addition, it is shown in Fig.~\ref{mixing_phiA} that the loop effects are
negligible for the mixing of the heavy Higgs bosons. Not shown are differences
between the different methods of evaluating the CP-mixing, i.e.\ whether the
parameters are evaluated at $M_{H^+}$ or at $m_t$, or whether $\mathcal M^2$ or
$(\mathcal M^2)^{\text{resum}}$ is used. The differences are on the same order
as the difference shown between the tree-level and the one-loop results, in
other words negligible.


\section{Conclusion}\label{se:conclusion}
\paragraph{}
In this paper, we have explored the Higgs sector of the CP-violating MSSM in a
mass scenario with heavy SUSY particles and light Higgs bosons using effective
field theory techniques. We matched the complex MSSM to a type-III 2HDM (where
both Higgs doublets couple to both the top and bottom quarks), and calculated
the complex threshold corrections to the 2HDM at one-loop level and the RGEs for
the 2HDM with complex parameters at two-loop level. Using these matching
conditions and evolving the parameters down to a low scale with these RGEs, we
resum contributions at NLL order. We explored the effect of including the
complex phases of the quartic couplings in the RGEs and found that in particular
the absolute values of $\lambda_5$ to $\lambda_7$ change substantially compared
to a scenario where only the absolute values (and signs) but not the phases are
included in the RGEs.

In order to calculate the pole masses of the Higgs bosons, we exploited
different methods:
\begin{enumerate}
\item[(a)] Calculation of the pole mass with the 2HDM with parameters at the scale of the mass of the charged Higgs boson $M_{H^+}$.
\item[(b)] Calculation of the pole mass with the 2HDM with parameters at the scale of the running top quark mass.
\item[(c)] Matching to the SM at the scale $M_{H^+}$ and calculation of the pole mass of the Higgs boson at the scale of the running top quark mass within the SM.
\item[(d)] Exploiting an approximation which resums the most important logarithms of
the form $\ln(\frac{M_{H^+}}{m_t})$ but still uses the full 2HDM for the calculation of the pole masses.
\end{enumerate}
The approximation (d) agrees well with the pure 2HDM result for small values of
$M_{H^+}$ and approaches the result where the SM is used as the low-scale effective
theory for larger $M_{H^+}$. Therefore, it can be used as an interpolation between
the two regimes.

We investigated several different scenarios to discuss the phase dependence of
both the masses of the Higgs bosons and the CP-violating components of the Higgs bosons.
All the masses of the Higgs bosons show a sizeable dependence on the common
phase $\varphi_A$ of $A_t$ and $A_b$, which can be of the order of several GeV
for low $\tan \beta$, in particular for the mass of the lightest Higgs boson.
The heaviest Higgs boson shows the least sensitivity to $\varphi_A$. The
dependence of the Higgs masses on the gluino phase is much weaker and of the
order of one GeV.

Additionally, we found that the size of the CP-odd component of the two heavy
Higgs bosons shows a negligible dependence on the mass of the charged Higgs
bosons, and that the heavy Higgs bosons interchange their CP-oddness when
varying $\varphi_A$. For the light Higgs boson, the size of the CP-odd component
decreases quickly with larger values of $M_{H^+}$ as has been discussed before
in e.g.\ Ref.~\cite{Li:2015yla}. Even though we find that the CP-odd admixture
in the considered scenario for $M_{H^+} = 500$ GeV is just below the expected
experimental reach according to the discussion in Ref.~\cite{Berge:2015nua}, it
is likely that this particular scenario is excluded by the experimental results
for the Higgs signal rates as well as of the measurement of electric dipole moments.
In order to find out whether there is a viable scenario with heavy SUSY partners
that leads to an observable size of a CP-odd component, further work is needed.

\subsection*{Acknowledgements}

We would like to thank Henning Bahl, Thomas Hahn, Sven Heinemeyer, Esben M\o
lgaard, Pietro Slavich, Florian Staub, Dominik St\"ockinger, and Georg Weiglein
for helpful discussions, as well as Joel Oredsson and Anders E. Thomsen for providing  their RGEs for comparison.
Our work is partially funded by the Danish National Research Foundation, grant
number DNRF90. Part of this work was supported by a STSM Grant from COST Action
CA16201 PARTICLEFACE.  In addition, H.R. was partly supported by the
German Federal Ministry for Education 
and Research (BMBF) under contract no.\ 05H18VFCA1 and partly funded by  the
Deutsche Forschungsgemeinschaft (DFG, German Research
Foundation)---project no.\ 442089526.
\appendix

\section{Yukawa Phases in the  Type-III 2HDM}
\paragraph{}
In a type-III 2HDM, complex phases in the Yukawa couplings cannot be rotated
away completely by absorbing the phases in the right-handed quark fields. To see
why this is true, consider the mass term for the top quark coming from
Eq.~\eqref{yukawas} after electroweak symmetry breaking
\be
\mathcal L_{\text{mass}}^{\text{top Yuk.}} =\frac{v}{\sqrt{2}}\left(h_t s_\beta + h'_t c_\beta\right)t t_c + \text{h.c.}. 
\ee
In order to make the top-quark mass real, the phase of the combination $\left(h_t
s_\beta + h'_t c_\beta\right)$ must be absorbed into the charge-conjugate quark
field. However, this phase will pop up again in the Yukawa couplings and will
not be cancelled by the phase of $h_t$ or $h'_t$. This means that, for example,
the top Yukawa term proportional to $h_t$ after the rotation now takes the form 
\be
\mathcal L^{\text{top Yuk.}}  = e^{-i\phi}h_t \epsilon_{ij} \Phi_2^i t_c Q^j + \text{h.c.}, 
\ee
where $\phi\equiv\arg(h_t s_\beta + h'_t c_\beta)$. These phases must be taken
into account when calculating the self-energy and tadpole corrections to the
Higgs mass matrix. However, we find that the effect is numerically negligble, as
the phase $\phi$ is small. 

\section{Tadpole Contributions}\label{tadpole}
In the following, we list the combinations of the tadpole contributions of the Higgs boson fields in the gauge basis $T_{\phi_1}$, $T_{\phi_2}$ and $T_{a_1}$ as they appear in the renormalized tadpole matrix $\hat{T}$ in Eqs.~\eqref{eq:det} and \eqref{mass_mat} and in the corresponding unrenormalized tadpole matrix $T$, 
\begin{align}
T_{11} &= \frac{c_{\beta} \Big[\left(c_{\beta}^2+2 s_{\beta}^2\right) T_{\phi_1}-c_{\beta} s_{\beta} T_{\phi_2}\Big]}{v},\\
T_{12} &= \frac{c_{\beta}^3 T_{\phi_2}+s_{\beta}^3 T_{\phi_1}}{v} = T_{21},\\
T_{13} &= 0 = T_{31},\\
T_{14} &= -\frac{T_{a_1}}{s_{\beta} v} = T_{41},\\
T_{22} &= \frac{s_{\beta} \Big[\left(2 c_{\beta}^2 +s_{\beta}^2 \right)T_{\phi_2}-c_{\beta} s_{\beta} T_{\phi_1}\Big]}{v},\\
T_{23} &= \frac{T_{a_1}}{s_{\beta} v} =T_{32},\\
T_{24} &= 0 = T_{42} ,\\
T_{33} &= T_{11}
,\\
T_{34} &= 
T_{12} = T_{43},\\
T_{44} &= T_{22}
.\\
\end{align}

\section{Conversion of the Vacuum Expectation Value}\label{se:vevconversion}

In Sect.~\ref{se:results}, the conversion of the vacuum expectation value is
explained briefly. In this appendix, for completeness, we list the explicit
conversion formulas.

The finite part of the counterterm to the on-shell vev
$\delta{v}^2_{\text{OS}-\mathrm{finite}}$ is given as
\begin{align}\label{eq:vevOSCT}
\delta{v}^2_{\text{OS}-\mathrm{finite}} = v^2_{\text{OS}}\left[\left(1-\frac{c_W^2}{s_W^2}\right)\frac{\delta M_W^2}{M_W^2} + \frac{c_W^2}{s_W^2}\frac{\delta M_Z^2}{M_Z^2} -\frac{\delta e^2}{e^2}\right]_{\text{finite}}
\end{align}
where $c_W^2 = 1 - s_W^2$. The W and the Z mass counterterm are chosen on-shell
via the one-loop pole mass definition while the counterterm for the electric
charge is fixed via the electron positron photon vertex in the Thomson limit and
are thus
\begin{align}
\delta M_V^2 = \text{Re} \Sigma_{VV}^\text{T}(M_V^2) \quad \text{with} \quad V = W,Z,\\
\delta e^2 = 2 e^2 \left[ \frac{1}{2} \left.\frac{\partial \text{Re}\Sigma_{\gamma\gamma} (k^2)}{\partial k^2}\right|_{k^2 = 0} - \frac{s_W}{c_W} \frac{\Sigma^{\text{T}}_{\gamma Z}(0)}{M_Z^2}\right]
\end{align}
where $\Sigma_{VV}^\text{T}$ is the transversal part of the W or the Z~boson
self energy at one-loop order, respectively. The one-loop photon self energy is
denoted by $\Sigma_{\gamma\gamma}$ and the transversal part of the one-loop
photon Z boson mixing as $\Sigma^{\text{T}}_{\gamma Z}$.

For $\Delta r$~\cite{Sirlin:1980prd}, we employed the one-loop result
\begin{align}\nonumber
\Delta r &= \frac{\delta e^2}{e^2} - \left(1 - \frac{c_W^2}{s_W^2}\right) \frac{\delta M_W^2}{M_W^2} - \frac{c_W^2}{s_W^2}\frac{\delta M_Z^2}{M_Z^2} +  \frac{2 \Sigma^{\text{T}}_{\gamma Z}(0)}{c_W s_W M_Z^2} + \frac{\Sigma^{\text{T}}_{WW}(0)}{M_W^2} \\&\quad + \frac{e^2}{32 \pi^2 s_W^4}\left[12 s_W^2 +( 7 - 4 s_W^2) \ln\left(\frac{M_W^2}{M_Z^2}\right) \right]
\end{align}
so that the complete one-loop conversion can be written as
\begin{align}
v_{\overline{\text{MS}}}^2 = v_{G_F}^2\left\{1 +  \frac{2 \Sigma^{\text{T}}_{\gamma Z}(0)}{c_W s_W M_Z^2} + \frac{\Sigma^{\text{T}}_{WW}(0)}{M_W^2} + \frac{e^2}{32 \pi^2 s_W^4}\left[12 s_W^2 +( 7 - 4 s_W^2) \ln\left(\frac{M_W^2}{M_Z^2}\right) \right]\right\}.
\end{align}


\clearpage
\bibliographystyle{JHEPmod}            
\bibliography{bibliography}

\providecommand{\href}[2]{#2}\begingroup\raggedright\begin{thebibliography}{100}

\bibitem{Aad:2012tfa}
{\bf ATLAS} Collaboration, G.~Aad et~al., {\it {Observation of a new particle
  in the search for the Standard Model Higgs boson with the ATLAS detector at
  the LHC}},  {\em Phys. Lett.} {\bf B716} (2012) 1--29,
  [\href{http://arxiv.org/abs/1207.7214}{{\tt arXiv:1207.7214}}].

\bibitem{Chatrchyan:2012xdj}
{\bf CMS} Collaboration, S.~Chatrchyan et~al., {\it {Observation of a new boson
  at a mass of 125 GeV with the CMS experiment at the LHC}},  {\em Phys. Lett.}
  {\bf B716} (2012) 30--61, [\href{http://arxiv.org/abs/1207.7235}{{\tt
  arXiv:1207.7235}}].

\bibitem{Aad:2015zhl}
{\bf ATLAS, CMS} Collaboration, G.~Aad et~al., {\it {Combined Measurement of
  the Higgs Boson Mass in $pp$ Collisions at $\sqrt{s}=7$ and 8 TeV with the
  ATLAS and CMS Experiments}},  {\em Phys. Rev. Lett.} {\bf 114} (2015) 191803,
  [\href{http://arxiv.org/abs/1503.07589}{{\tt arXiv:1503.07589}}].

\bibitem{Haber:1990aw}
H.~E. Haber and R.~Hempfling, {\it {Can the mass of the lightest Higgs boson of
  the minimal supersymmetric model be larger than m(Z)?}},  {\em Phys. Rev.
  Lett.} {\bf 66} (1991) 1815--1818.

\bibitem{Ellis:1990nz}
J.~R. Ellis, G.~Ridolfi, and F.~Zwirner, {\it {Radiative corrections to the
  masses of supersymmetric Higgs bosons}},  {\em Phys. Lett.} {\bf B257} (1991)
  83--91.

\bibitem{Okada:1990vk}
Y.~Okada, M.~Yamaguchi, and T.~Yanagida, {\it {Upper bound of the lightest
  Higgs boson mass in the minimal supersymmetric standard model}},  {\em Prog.
  Theor. Phys.} {\bf 85} (1991) 1--6.

\bibitem{Ellis:1991zd}
J.~R. Ellis, G.~Ridolfi, and F.~Zwirner, {\it {On radiative corrections to
  supersymmetric Higgs boson masses and their implications for LEP searches}},
  {\em Phys. Lett.} {\bf B262} (1991) 477--484.

\bibitem{Chankowski:1991md}
P.~H. Chankowski, S.~Pokorski, and J.~Rosiek, {\it {Charged and neutral
  supersymmetric Higgs boson masses: Complete one loop analysis}},  {\em Phys.
  Lett.} {\bf B274} (1992) 191--198.

\bibitem{Brignole:1992uf}
A.~Brignole, {\it {Radiative corrections to the supersymmetric neutral Higgs
  boson masses}},  {\em Phys. Lett.} {\bf B281} (1992) 284--294.

\bibitem{Chankowski:1992er}
P.~H. Chankowski, S.~Pokorski, and J.~Rosiek, {\it {Complete on-shell
  renormalization scheme for the minimal supersymmetric Higgs sector}},  {\em
  Nucl. Phys.} {\bf B423} (1994) 437--496,
  [\href{http://arxiv.org/abs/hep-ph/9303309}{{\tt hep-ph/9303309}}].

\bibitem{Dabelstein:1994hb}
A.~Dabelstein, {\it {The One loop renormalization of the MSSM Higgs sector and
  its application to the neutral scalar Higgs masses}},  {\em Z. Phys.} {\bf
  C67} (1995) 495--512, [\href{http://arxiv.org/abs/hep-ph/9409375}{{\tt
  hep-ph/9409375}}].

\bibitem{Pierce:1996zz}
D.~M. Pierce, J.~A. Bagger, K.~T. Matchev, and R.-j. Zhang, {\it {Precision
  corrections in the minimal supersymmetric standard model}},  {\em Nucl.
  Phys.} {\bf B491} (1997) 3--67,
  [\href{http://arxiv.org/abs/hep-ph/9606211}{{\tt hep-ph/9606211}}].

\bibitem{Frank:2006yh}
M.~Frank, et~al., {\it {The Higgs boson masses and mixings of the complex MSSM
  in the Feynman-diagrammatic approach}},  {\em JHEP} {\bf 02} (2007) 047,
  [\href{http://arxiv.org/abs/hep-ph/0611326}{{\tt hep-ph/0611326}}].

\bibitem{Heinemeyer:1998jw}
S.~Heinemeyer, W.~Hollik, and G.~Weiglein, {\it {QCD corrections to the masses
  of the neutral CP - even Higgs bosons in the MSSM}},  {\em Phys. Rev.} {\bf
  D58} (1998) 091701, [\href{http://arxiv.org/abs/hep-ph/9803277}{{\tt
  hep-ph/9803277}}].

\bibitem{Heinemeyer:1998kz}
S.~Heinemeyer, W.~Hollik, and G.~Weiglein, {\it {Precise prediction for the
  mass of the lightest Higgs boson in the MSSM}},  {\em Phys. Lett.} {\bf B440}
  (1998) 296--304, [\href{http://arxiv.org/abs/hep-ph/9807423}{{\tt
  hep-ph/9807423}}].

\bibitem{Zhang:1998bm}
R.-J. Zhang, {\it {Two loop effective potential calculation of the lightest CP
  even Higgs boson mass in the MSSM}},  {\em Phys. Lett.} {\bf B447} (1999)
  89--97, [\href{http://arxiv.org/abs/hep-ph/9808299}{{\tt hep-ph/9808299}}].

\bibitem{Heinemeyer:1998np}
S.~Heinemeyer, W.~Hollik, and G.~Weiglein, {\it {The Masses of the neutral CP -
  even Higgs bosons in the MSSM: Accurate analysis at the two loop level}},
  {\em Eur. Phys. J.} {\bf C9} (1999) 343--366,
  [\href{http://arxiv.org/abs/hep-ph/9812472}{{\tt hep-ph/9812472}}].

\bibitem{Espinosa:1999zm}
J.~R. Espinosa and R.-J. Zhang, {\it {MSSM lightest CP even Higgs boson mass to
  $O(\alpha_s \alpha_t)$: The effective potential approach}},  {\em JHEP} {\bf
  03} (2000) 026, [\href{http://arxiv.org/abs/hep-ph/9912236}{{\tt
  hep-ph/9912236}}].

\bibitem{Espinosa:2000df}
J.~R. Espinosa and R.-J. Zhang, {\it {Complete two loop dominant corrections to
  the mass of the lightest CP even Higgs boson in the minimal supersymmetric
  standard model}},  {\em Nucl. Phys.} {\bf B586} (2000) 3--38,
  [\href{http://arxiv.org/abs/hep-ph/0003246}{{\tt hep-ph/0003246}}].

\bibitem{Degrassi:2001yf}
G.~Degrassi, P.~Slavich, and F.~Zwirner, {\it {On the neutral Higgs boson
  masses in the MSSM for arbitrary stop mixing}},  {\em Nucl. Phys.} {\bf B611}
  (2001) 403--422, [\href{http://arxiv.org/abs/hep-ph/0105096}{{\tt
  hep-ph/0105096}}].

\bibitem{Brignole:2001jy}
A.~Brignole, G.~Degrassi, P.~Slavich, and F.~Zwirner, {\it {On the
  $O(\alpha_t^2$) two loop corrections to the neutral Higgs boson masses in the
  MSSM}},  {\em Nucl. Phys.} {\bf B631} (2002) 195--218,
  [\href{http://arxiv.org/abs/hep-ph/0112177}{{\tt hep-ph/0112177}}].

\bibitem{Brignole:2002bz}
A.~Brignole, G.~Degrassi, P.~Slavich, and F.~Zwirner, {\it {On the two loop
  sbottom corrections to the neutral Higgs boson masses in the MSSM}},  {\em
  Nucl. Phys.} {\bf B643} (2002) 79--92,
  [\href{http://arxiv.org/abs/hep-ph/0206101}{{\tt hep-ph/0206101}}].

\bibitem{Martin:2002iu}
S.~P. Martin, {\it {Two loop effective potential for the minimal supersymmetric
  standard model}},  {\em Phys. Rev.} {\bf D66} (2002) 096001,
  [\href{http://arxiv.org/abs/hep-ph/0206136}{{\tt hep-ph/0206136}}].

\bibitem{Martin:2002wn}
S.~P. Martin, {\it {Complete two loop effective potential approximation to the
  lightest Higgs scalar boson mass in supersymmetry}},  {\em Phys. Rev.} {\bf
  D67} (2003) 095012, [\href{http://arxiv.org/abs/hep-ph/0211366}{{\tt
  hep-ph/0211366}}].

\bibitem{Dedes:2002dy}
A.~Dedes and P.~Slavich, {\it {Two loop corrections to radiative electroweak
  symmetry breaking in the MSSM}},  {\em Nucl. Phys.} {\bf B657} (2003)
  333--354, [\href{http://arxiv.org/abs/hep-ph/0212132}{{\tt hep-ph/0212132}}].

\bibitem{Dedes:2003km}
A.~Dedes, G.~Degrassi, and P.~Slavich, {\it {On the two loop Yukawa corrections
  to the MSSM Higgs boson masses at large $\tan\beta$}},  {\em Nucl. Phys.}
  {\bf B672} (2003) 144--162, [\href{http://arxiv.org/abs/hep-ph/0305127}{{\tt
  hep-ph/0305127}}].

\bibitem{Martin:2004kr}
S.~P. Martin, {\it {Strong and Yukawa two-loop contributions to Higgs scalar
  boson self-energies and pole masses in supersymmetry}},  {\em Phys. Rev.}
  {\bf D71} (2005) 016012, [\href{http://arxiv.org/abs/hep-ph/0405022}{{\tt
  hep-ph/0405022}}].

\bibitem{Allanach:2004rh}
B.~Allanach, A.~Djouadi, J.~Kneur, W.~Porod, and P.~Slavich, {\it {Precise
  determination of the neutral Higgs boson masses in the MSSM}},  {\em JHEP}
  {\bf 0409} (2004) 044, [\href{http://arxiv.org/abs/hep-ph/0406166}{{\tt
  hep-ph/0406166}}].

\bibitem{Heinemeyer:2004xw}
S.~Heinemeyer, W.~Hollik, H.~Rzehak, and G.~Weiglein, {\it {High-precision
  predictions for the MSSM Higgs sector at $O(\alpha_b \alpha_s)$}},  {\em Eur.
  Phys. J.} {\bf C39} (2005) 465--481,
  [\href{http://arxiv.org/abs/hep-ph/0411114}{{\tt hep-ph/0411114}}].

\bibitem{Martin:2005eg}
S.~P. Martin, {\it {Two-loop scalar self-energies and pole masses in a general
  renormalizable theory with massless gauge bosons}},  {\em Phys. Rev.} {\bf
  D71} (2005) 116004, [\href{http://arxiv.org/abs/hep-ph/0502168}{{\tt
  hep-ph/0502168}}].

\bibitem{Heinemeyer:2007aq}
S.~Heinemeyer, W.~Hollik, H.~Rzehak, and G.~Weiglein, {\it {The Higgs sector of
  the complex MSSM at two-loop order: QCD contributions}},  {\em Phys.Lett.}
  {\bf B652} (2007) 300--309, [\href{http://arxiv.org/abs/0705.0746}{{\tt
  arXiv:0705.0746}}].

\bibitem{Borowka:2014wla}
S.~Borowka, T.~Hahn, S.~Heinemeyer, G.~Heinrich, and W.~Hollik, {\it
  {Momentum-dependent two-loop QCD corrections to the neutral Higgs-boson
  masses in the MSSM}},  {\em Eur. Phys. J.} {\bf C74} (2014), no.~8 2994,
  [\href{http://arxiv.org/abs/1404.7074}{{\tt arXiv:1404.7074}}].

\bibitem{Degrassi:2014pfa}
G.~Degrassi, S.~Di~Vita, and P.~Slavich, {\it {Two-loop QCD corrections to the
  MSSM Higgs masses beyond the effective-potential approximation}},  {\em Eur.
  Phys. J.} {\bf C75} (2015), no.~2 61,
  [\href{http://arxiv.org/abs/1410.3432}{{\tt arXiv:1410.3432}}].

\bibitem{Hollik:2014wea}
W.~Hollik and S.~Pa{\ss}ehr, {\it {Two-loop top-Yukawa-coupling corrections to
  the Higgs boson masses in the complex MSSM}},  {\em Phys. Lett.} {\bf B733}
  (2014) 144--150, [\href{http://arxiv.org/abs/1401.8275}{{\tt
  arXiv:1401.8275}}].

\bibitem{Hollik:2014bua}
W.~Hollik and S.~Pa{\ss}ehr, {\it {Higgs boson masses and mixings in the
  complex MSSM with two-loop top-Yukawa-coupling corrections}},  {\em JHEP}
  {\bf 10} (2014) 171, [\href{http://arxiv.org/abs/1409.1687}{{\tt
  arXiv:1409.1687}}].

\bibitem{Hollik:2015ema}
W.~Hollik and S.~Pa{\ss}ehr, {\it {Two-loop top-Yukawa-coupling corrections to
  the charged Higgs-boson mass in the MSSM}},  {\em Eur. Phys. J.} {\bf C75}
  (2015), no.~7 336, [\href{http://arxiv.org/abs/1502.02394}{{\tt
  arXiv:1502.02394}}].

\bibitem{Borowka:2015ura}
S.~Borowka, T.~Hahn, S.~Heinemeyer, G.~Heinrich, and W.~Hollik, {\it
  {Renormalization scheme dependence of the two-loop QCD corrections to the
  neutral Higgs-boson masses in the MSSM}},  {\em Eur. Phys. J.} {\bf C75}
  (2015), no.~9 424, [\href{http://arxiv.org/abs/1505.03133}{{\tt
  arXiv:1505.03133}}].

\bibitem{Goodsell:2016udb}
M.~D. Goodsell and F.~Staub, {\it {The Higgs mass in the CP violating MSSM,
  NMSSM, and beyond}},  {\em Eur. Phys. J.} {\bf C77} (2017), no.~1 46,
  [\href{http://arxiv.org/abs/1604.05335}{{\tt arXiv:1604.05335}}].

\bibitem{Passehr:2017ufr}
S.~Pa{\ss}ehr and G.~Weiglein, {\it {Two-loop top and bottom Yukawa corrections
  to the Higgs-boson masses in the complex MSSM}},  {\em Eur. Phys. J.} {\bf
  C78} (2018), no.~3 222, [\href{http://arxiv.org/abs/1705.07909}{{\tt
  arXiv:1705.07909}}].

\bibitem{Borowka:2018anu}
S.~Borowka, S.~Pa{\ss}ehr, and G.~Weiglein, {\it {Complete two-loop QCD
  contributions to the lightest Higgs-boson mass in the MSSM with complex
  parameters}},  {\em Eur. Phys. J.} {\bf C78} (2018), no.~7 576,
  [\href{http://arxiv.org/abs/1802.09886}{{\tt arXiv:1802.09886}}].

\bibitem{Martin:2007pg}
S.~P. Martin, {\it {Three-loop corrections to the lightest Higgs scalar boson
  mass in supersymmetry}},  {\em Phys. Rev.} {\bf D75} (2007) 055005,
  [\href{http://arxiv.org/abs/hep-ph/0701051}{{\tt hep-ph/0701051}}].

\bibitem{Harlander:2008ju}
R.~Harlander, P.~Kant, L.~Mihaila, and M.~Steinhauser, {\it {Higgs boson mass
  in supersymmetry to three loops}},  {\em Phys.Rev.Lett.} {\bf 100} (2008)
  191602, [\href{http://arxiv.org/abs/0803.0672}{{\tt arXiv:0803.0672}}].

\bibitem{Kant:2010tf}
P.~Kant, R.~V. Harlander, L.~Mihaila, and M.~Steinhauser, {\it {Light MSSM
  Higgs boson mass to three-loop accuracy}},  {\em JHEP} {\bf 08} (2010) 104,
  [\href{http://arxiv.org/abs/1005.5709}{{\tt arXiv:1005.5709}}].

\bibitem{Harlander:2017kuc}
R.~V. Harlander, J.~Klappert, and A.~Voigt, {\it {Higgs mass prediction in the
  MSSM at three-loop level in a pure $\overline{{\text {DR}}}$ context}},  {\em
  Eur. Phys. J.} {\bf C77} (2017), no.~12 814,
  [\href{http://arxiv.org/abs/1708.05720}{{\tt arXiv:1708.05720}}].

\bibitem{Stockinger:2018oxe}
D.~St{\"o}ckinger and J.~Unger, {\it {Three-loop MSSM Higgs-boson mass
  predictions and regularization by dimensional reduction}},  {\em Nucl. Phys.}
  {\bf B935} (2018) 1--16, [\href{http://arxiv.org/abs/1804.05619}{{\tt
  arXiv:1804.05619}}].

\bibitem{R.:2019ply}
A.~R. Fazio and E.~A. Reyes~R., {\it {The Lightest Higgs Boson Mass of the MSSM
  at Three-Loop Accuracy}},  {\em Nucl. Phys.} {\bf B942} (2019) 164--183,
  [\href{http://arxiv.org/abs/1901.03651}{{\tt arXiv:1901.03651}}].

\bibitem{R.:2019irs}
E.~A. Reyes~R. and A.~R. Fazio, {\it {Comparison of the EFT Hybrid and
  Three-Loop Fixed-Order Calculations of the Lightest MSSM Higgs Boson Mass}},
  \href{http://arxiv.org/abs/1908.00693}{{\tt arXiv:1908.00693}}.

\bibitem{Degrassi:2002fi}
G.~Degrassi, S.~Heinemeyer, W.~Hollik, P.~Slavich, and G.~Weiglein, {\it
  {Towards high precision predictions for the MSSM Higgs sector}},  {\em Eur.
  Phys. J.} {\bf C28} (2003) 133--143,
  [\href{http://arxiv.org/abs/hep-ph/0212020}{{\tt hep-ph/0212020}}].

\bibitem{Vega:2015fna}
J.~P. Vega and G.~Villadoro, {\it {SusyHD: Higgs mass determination in
  supersymmetry}},  {\em JHEP} {\bf 07} (2015) 159,
  [\href{http://arxiv.org/abs/1504.05200}{{\tt arXiv:1504.05200}}].

\bibitem{Bahl:2017aev}
H.~Bahl, S.~Heinemeyer, W.~Hollik, and G.~Weiglein, {\it {Reconciling EFT and
  hybrid calculations of the light MSSM Higgs-boson mass}},  {\em Eur. Phys.
  J.} {\bf C78} (2018), no.~1 57, [\href{http://arxiv.org/abs/1706.00346}{{\tt
  arXiv:1706.00346}}].

\bibitem{Allanach:2018fif}
B.~C. Allanach and A.~Voigt, {\it {Uncertainties in the Lightest $CP$ Even
  Higgs Boson Mass Prediction in the Minimal Supersymmetric Standard Model:
  Fixed Order Versus Effective Field Theory Prediction}},  {\em Eur. Phys. J.}
  {\bf C78} (2018), no.~7 573, [\href{http://arxiv.org/abs/1804.09410}{{\tt
  arXiv:1804.09410}}].

\bibitem{Bahl:2019hmm}
H.~Bahl, S.~Heinemeyer, W.~Hollik, and G.~Weiglein, {\it {Theoretical
  uncertainties in the MSSM Higgs boson mass calculation}},  {\em Eur. Phys. J.
  C} {\bf 80} (2020), no.~6 497, [\href{http://arxiv.org/abs/1912.04199}{{\tt
  arXiv:1912.04199}}].

\bibitem{Slavich:2020zjv}
P.~Slavich et~al., {\it {Higgs-mass predictions in the MSSM and beyond}},  {\em
  Eur. Phys. J. C} {\bf 81} (2021), no.~5 450,
  [\href{http://arxiv.org/abs/2012.15629}{{\tt arXiv:2012.15629}}].

\bibitem{Domingo:2021kud}
F.~Domingo and S.~Pa\ss{}ehr, {\it {Fighting off field dependence in MSSM
  Higgs-mass corrections of order $\alpha _t\,\alpha _s$ and $\alpha _t^2$}},
  {\em Eur. Phys. J. C} {\bf 81} (2021), no.~7 661,
  [\href{http://arxiv.org/abs/2105.01139}{{\tt arXiv:2105.01139}}].

\bibitem{R:2021bml}
E.~A.~R. R. and R.~Fazio, {\it {High-Precision Calculations of the Higgs Boson
  Mass}},  {\em Particles} {\bf 5} (2022), no.~1 53--73,
  [\href{http://arxiv.org/abs/2112.15295}{{\tt arXiv:2112.15295}}].

\bibitem{Barbieri:1990ja}
R.~Barbieri, M.~Frigeni, and F.~Caravaglios, {\it {The Supersymmetric Higgs for
  heavy superpartners}},  {\em Phys. Lett.} {\bf B258} (1991) 167--170.

\bibitem{Okada:1990gg}
Y.~Okada, M.~Yamaguchi, and T.~Yanagida, {\it {Renormalization group analysis
  on the Higgs mass in the softly broken supersymmetric standard model}},  {\em
  Phys. Lett.} {\bf B262} (1991) 54--58.

\bibitem{Kodaira:1993yt}
J.~Kodaira, Y.~Yasui, and K.~Sasaki, {\it {The Mass of the lightest
  supersymmetric Higgs boson beyond the leading logarithm approximation}},
  {\em Phys. Rev.} {\bf D50} (1994) 7035--7041,
  [\href{http://arxiv.org/abs/hep-ph/9311366}{{\tt hep-ph/9311366}}].

\bibitem{Hempfling:1993qq}
R.~Hempfling and A.~H. Hoang, {\it {Two loop radiative corrections to the upper
  limit of the lightest Higgs boson mass in the minimal supersymmetric model}},
   {\em Phys. Lett.} {\bf B331} (1994) 99--106,
  [\href{http://arxiv.org/abs/hep-ph/9401219}{{\tt hep-ph/9401219}}].

\bibitem{Casas:1994us}
J.~A. Casas, J.~R. Espinosa, M.~Quiros, and A.~Riotto, {\it {The Lightest Higgs
  boson mass in the minimal supersymmetric standard model}},  {\em Nucl. Phys.}
  {\bf B436} (1995) 3--29, [\href{http://arxiv.org/abs/hep-ph/9407389}{{\tt
  hep-ph/9407389}}]. [Erratum: Nucl. Phys.B439,466(1995)].

\bibitem{Haber:1996fp}
H.~E. Haber, R.~Hempfling, and A.~H. Hoang, {\it {Approximating the radiatively
  corrected Higgs mass in the minimal supersymmetric model}},  {\em Z. Phys.}
  {\bf C75} (1997) 539--554, [\href{http://arxiv.org/abs/hep-ph/9609331}{{\tt
  hep-ph/9609331}}].

\bibitem{Espinosa:1991fc}
J.~R. Espinosa and M.~Quiros, {\it {Two loop radiative corrections to the mass
  of the lightest Higgs boson in supersymmetric standard models}},  {\em Phys.
  Lett.} {\bf B266} (1991) 389--396.

\bibitem{Sasaki:1991qu}
K.~Sasaki, M.~Carena, and C.~E.~M. Wagner, {\it {Renormalization group analysis
  of the Higgs sector in the minimal supersymmetric standard model}},  {\em
  Nucl. Phys.} {\bf B381} (1992) 66--86.

\bibitem{Chankowski:1992ek}
P.~H. Chankowski, S.~Pokorski, and J.~Rosiek, {\it {Is the lightest
  supersymmetric Higgs Boson distinguishable from the minimal standard model
  one?}},  {\em Phys. Lett.} {\bf B281} (1992) 100--105.

\bibitem{Haber:1993an}
H.~E. Haber and R.~Hempfling, {\it {The renormalization group improved Higgs
  sector of the minimal supersymmetric model}},  {\em Phys. Rev.} {\bf D48}
  (1993) 4280--4309, [\href{http://arxiv.org/abs/hep-ph/9307201}{{\tt
  hep-ph/9307201}}].

\bibitem{Carena:1995bx}
M.~Carena, J.~R. Espinosa, M.~Quiros, and C.~E.~M. Wagner, {\it {Analytical
  expressions for radiatively corrected Higgs masses and couplings in the
  MSSM}},  {\em Phys. Lett.} {\bf B355} (1995) 209--221,
  [\href{http://arxiv.org/abs/hep-ph/9504316}{{\tt hep-ph/9504316}}].

\bibitem{Espinosa:2001mm}
J.~R. Espinosa and I.~Navarro, {\it {Radiative corrections to the Higgs boson
  mass for a hierarchical stop spectrum}},  {\em Nucl. Phys.} {\bf B615} (2001)
  82--116, [\href{http://arxiv.org/abs/hep-ph/0104047}{{\tt hep-ph/0104047}}].

\bibitem{Pilaftsis:1999qt}
A.~Pilaftsis and C.~E.~M. Wagner, {\it {Higgs bosons in the minimal
  supersymmetric standard model with explicit CP violation}},  {\em Nucl.
  Phys.} {\bf B553} (1999) 3--42,
  [\href{http://arxiv.org/abs/hep-ph/9902371}{{\tt hep-ph/9902371}}].

\bibitem{Carena:2000yi}
M.~Carena, J.~R. Ellis, A.~Pilaftsis, and C.~E.~M. Wagner, {\it
  {Renormalization group improved effective potential for the MSSM Higgs sector
  with explicit CP violation}},  {\em Nucl. Phys.} {\bf B586} (2000) 92--140,
  [\href{http://arxiv.org/abs/hep-ph/0003180}{{\tt hep-ph/0003180}}].

\bibitem{Draper:2013oza}
P.~Draper, G.~Lee, and C.~E.~M. Wagner, {\it {Precise estimates of the Higgs
  mass in heavy supersymmetry}},  {\em Phys. Rev.} {\bf D89} (2014), no.~5
  055023, [\href{http://arxiv.org/abs/1312.5743}{{\tt arXiv:1312.5743}}].

\bibitem{Lee:2015uza}
G.~Lee and C.~E.~M. Wagner, {\it {Higgs bosons in heavy supersymmetry with an
  intermediate m$_A$}},  {\em Phys. Rev.} {\bf D92} (2015), no.~7 075032,
  [\href{http://arxiv.org/abs/1508.00576}{{\tt arXiv:1508.00576}}].

\bibitem{Bagnaschi:2015pwa}
E.~Bagnaschi, F.~Br{\"u}mmer, W.~Buchm{\"u}ller, A.~Voigt, and G.~Weiglein,
  {\it {Vacuum stability and supersymmetry at high scales with two Higgs
  doublets}},  {\em JHEP} {\bf 03} (2016) 158,
  [\href{http://arxiv.org/abs/1512.07761}{{\tt arXiv:1512.07761}}].

\bibitem{Gorbahn:2009pp}
M.~Gorbahn, S.~J{\"a}ger, U.~Nierste, and S.~Trine, {\it {The supersymmetric
  Higgs sector and $B-\bar{B}$ mixing for large tan $\beta$}},  {\em Phys.
  Rev.} {\bf D84} (2011) 034030, [\href{http://arxiv.org/abs/0901.2065}{{\tt
  arXiv:0901.2065}}].

\bibitem{Bagnaschi:2014rsa}
E.~Bagnaschi, G.~F. Giudice, P.~Slavich, and A.~Strumia, {\it {Higgs Mass and
  Unnatural Supersymmetry}},  {\em JHEP} {\bf 09} (2014) 092,
  [\href{http://arxiv.org/abs/1407.4081}{{\tt arXiv:1407.4081}}].

\bibitem{Bagnaschi:2017xid}
E.~Bagnaschi, J.~Pardo~Vega, and P.~Slavich, {\it {Improved determination of
  the Higgs mass in the MSSM with heavy superpartners}},  {\em Eur. Phys. J.}
  {\bf C77} (2017), no.~5 334, [\href{http://arxiv.org/abs/1703.08166}{{\tt
  arXiv:1703.08166}}].

\bibitem{Bagnaschi:2019esc}
E.~Bagnaschi, G.~Degrassi, S.~Pa{\ss}ehr, and P.~Slavich, {\it {Full two-loop
  QCD corrections to the Higgs mass in the MSSM with heavy superpartners}},
  \href{http://arxiv.org/abs/1908.01670}{{\tt arXiv:1908.01670}}.

\bibitem{Wells:2017vla}
J.~D. Wells and Z.~Zhang, {\it {Effective field theory approach to trans-TeV
  supersymmetry: covariant matching, Yukawa unification and Higgs couplings}},
  {\em JHEP} {\bf 05} (2018) 182, [\href{http://arxiv.org/abs/1711.04774}{{\tt
  arXiv:1711.04774}}].

\bibitem{Harlander:2018yhj}
R.~V. Harlander, J.~Klappert, A.~D. Ochoa~Franco, and A.~Voigt, {\it {The light
  CP-even MSSM Higgs mass resummed to fourth logarithmic order}},  {\em Eur.
  Phys. J.} {\bf C78} (2018), no.~10 874,
  [\href{http://arxiv.org/abs/1807.03509}{{\tt arXiv:1807.03509}}].

\bibitem{Hahn:2013ria}
T.~Hahn, S.~Heinemeyer, W.~Hollik, H.~Rzehak, and G.~Weiglein, {\it
  {High-precision predictions for the light CP-even Higgs boson mass of the
  Minimal Supersymmetric Standard Model}},  {\em Phys. Rev. Lett.} {\bf 112}
  (2014), no.~14 141801, [\href{http://arxiv.org/abs/1312.4937}{{\tt
  arXiv:1312.4937}}].

\bibitem{Bahl:2016brp}
H.~Bahl and W.~Hollik, {\it {Precise prediction for the light MSSM Higgs boson
  mass combining effective field theory and fixed-order calculations}},  {\em
  Eur. Phys. J.} {\bf C76} (2016), no.~9 499,
  [\href{http://arxiv.org/abs/1608.01880}{{\tt arXiv:1608.01880}}].

\bibitem{Bahl:2018jom}
H.~Bahl and W.~Hollik, {\it {Precise prediction of the MSSM Higgs boson masses
  for low M$_{A}$}},  {\em JHEP} {\bf 07} (2018) 182,
  [\href{http://arxiv.org/abs/1805.00867}{{\tt arXiv:1805.00867}}].

\bibitem{Bahl:2019wzx}
H.~Bahl, I.~Sobolev, and G.~Weiglein, {\it {Precise prediction for the mass of
  the light MSSM Higgs boson for the case of a heavy gluino}},  {\em Phys.
  Lett. B} {\bf 808} (2020) 135644,
  [\href{http://arxiv.org/abs/1912.10002}{{\tt arXiv:1912.10002}}].

\bibitem{Bahl:2020tuq}
H.~Bahl, I.~Sobolev, and G.~Weiglein, {\it {The light MSSM Higgs boson mass for
  large $\tan \beta $ and complex input parameters}},  {\em Eur. Phys. J. C}
  {\bf 80} (2020), no.~11 1063, [\href{http://arxiv.org/abs/2009.07572}{{\tt
  arXiv:2009.07572}}].

\bibitem{Heinemeyer:1998yj}
S.~Heinemeyer, W.~Hollik, and G.~Weiglein, {\it {FeynHiggs: A Program for the
  calculation of the masses of the neutral CP even Higgs bosons in the MSSM}},
  {\em Comput. Phys. Commun.} {\bf 124} (2000) 76--89,
  [\href{http://arxiv.org/abs/hep-ph/9812320}{{\tt hep-ph/9812320}}].

\bibitem{Bahl:2018qog}
H.~Bahl, et~al., {\it {Precision calculations in the MSSM Higgs-boson sector
  with FeynHiggs 2.14}},  \href{http://arxiv.org/abs/1811.09073}{{\tt
  arXiv:1811.09073}}.

\bibitem{Athron:2014yba}
P.~Athron, J.-h. Park, D.~St{\"o}ckinger, and A.~Voigt, {\it {FlexibleSUSY?A
  spectrum generator generator for supersymmetric models}},  {\em Comput. Phys.
  Commun.} {\bf 190} (2015) 139--172,
  [\href{http://arxiv.org/abs/1406.2319}{{\tt arXiv:1406.2319}}].

\bibitem{Athron:2017fvs}
P.~Athron, et~al., {\it {FlexibleSUSY 2.0: Extensions to investigate the
  phenomenology of SUSY and non-SUSY models}},  {\em Comput. Phys. Commun.}
  {\bf 230} (2018) 145--217, [\href{http://arxiv.org/abs/1710.03760}{{\tt
  arXiv:1710.03760}}].

\bibitem{Staub:2009bi}
F.~Staub, {\it {From Superpotential to Model Files for FeynArts and
  CalcHep/CompHep}},  {\em Comput. Phys. Commun.} {\bf 181} (2010) 1077--1086,
  [\href{http://arxiv.org/abs/0909.2863}{{\tt arXiv:0909.2863}}].

\bibitem{Staub:2010jh}
F.~Staub, {\it {Automatic Calculation of supersymmetric Renormalization Group
  Equations and Self Energies}},  {\em Comput. Phys. Commun.} {\bf 182} (2011)
  808--833, [\href{http://arxiv.org/abs/1002.0840}{{\tt arXiv:1002.0840}}].

\bibitem{Staub:2012pb}
F.~Staub, {\it {SARAH 3.2: Dirac Gauginos, UFO output, and more}},  {\em
  Comput. Phys. Commun.} {\bf 184} (2013) 1792--1809,
  [\href{http://arxiv.org/abs/1207.0906}{{\tt arXiv:1207.0906}}].

\bibitem{Staub:2013tta}
F.~Staub, {\it {SARAH 4 : A tool for (not only SUSY) model builders}},  {\em
  Comput. Phys. Commun.} {\bf 185} (2014) 1773--1790,
  [\href{http://arxiv.org/abs/1309.7223}{{\tt arXiv:1309.7223}}].

\bibitem{Porod:2003um}
W.~Porod, {\it {SPheno, a program for calculating supersymmetric spectra, SUSY
  particle decays and SUSY particle production at e+ e- colliders}},  {\em
  Comput. Phys. Commun.} {\bf 153} (2003) 275--315,
  [\href{http://arxiv.org/abs/hep-ph/0301101}{{\tt hep-ph/0301101}}].

\bibitem{Porod:2011nf}
W.~Porod and F.~Staub, {\it {SPheno 3.1: Extensions including flavour,
  CP-phases and models beyond the MSSM}},  {\em Comput. Phys. Commun.} {\bf
  183} (2012) 2458--2469, [\href{http://arxiv.org/abs/1104.1573}{{\tt
  arXiv:1104.1573}}].

\bibitem{Athron:2016fuq}
P.~Athron, J.-h. Park, T.~Steudtner, D.~St{\"o}ckinger, and A.~Voigt, {\it
  {Precise Higgs mass calculations in (non-)minimal supersymmetry at both high
  and low scales}},  {\em JHEP} {\bf 01} (2017) 079,
  [\href{http://arxiv.org/abs/1609.00371}{{\tt arXiv:1609.00371}}].

\bibitem{Staub:2017jnp}
F.~Staub and W.~Porod, {\it {Improved predictions for intermediate and heavy
  Supersymmetry in the MSSM and beyond}},  {\em Eur. Phys. J.} {\bf C77}
  (2017), no.~5 338, [\href{http://arxiv.org/abs/1703.03267}{{\tt
  arXiv:1703.03267}}].

\bibitem{Harlander:2019dge}
R.~V. Harlander, J.~Klappert, and A.~Voigt, {\it {The light CP-even MSSM Higgs
  mass including N$^\mathbf {3}$LO+N$^\mathbf {3}$LL QCD corrections}},  {\em
  Eur. Phys. J. C} {\bf 80} (2020), no.~3 186,
  [\href{http://arxiv.org/abs/1910.03595}{{\tt arXiv:1910.03595}}].

\bibitem{Bahl:2020mjy}
H.~Bahl, N.~Murphy, and H.~Rzehak, {\it {Hybrid calculation of the MSSM Higgs
  boson masses using the complex THDM as EFT}},  {\em Eur. Phys. J. C} {\bf 81}
  (2021), no.~2 128, [\href{http://arxiv.org/abs/2010.04711}{{\tt
  arXiv:2010.04711}}].

\bibitem{Bahl:2020jaq}
H.~Bahl and I.~Sobolev, {\it {Two-loop matching of renormalizable operators:
  general considerations and applications}},  {\em JHEP} {\bf 03} (2021) 286,
  [\href{http://arxiv.org/abs/2010.01989}{{\tt arXiv:2010.01989}}].

\bibitem{Arbey:2014msa}
A.~Arbey, J.~Ellis, R.~M. Godbole, and F.~Mahmoudi, {\it {Exploring CP
  Violation in the MSSM}},  {\em Eur. Phys. J.} {\bf C75} (2015), no.~2 85,
  [\href{http://arxiv.org/abs/1410.4824}{{\tt arXiv:1410.4824}}].

\bibitem{Li:2015yla}
B.~Li and C.~E.~M. Wagner, {\it {CP-odd component of the lightest neutral Higgs
  boson in the MSSM}},  {\em Phys. Rev.} {\bf D91} (2015) 095019,
  [\href{http://arxiv.org/abs/1502.02210}{{\tt arXiv:1502.02210}}].

\bibitem{Carena:2015uoe}
M.~Carena, J.~Ellis, J.~S. Lee, A.~Pilaftsis, and C.~E.~M. Wagner, {\it {CP
  Violation in Heavy MSSM Higgs Scenarios}},  {\em JHEP} {\bf 02} (2016) 123,
  [\href{http://arxiv.org/abs/1512.00437}{{\tt arXiv:1512.00437}}].

\bibitem{Kobayashi:1973fv}
M.~Kobayashi and T.~Maskawa, {\it {CP Violation in the Renormalizable Theory of
  Weak Interaction}},  {\em Prog. Theor. Phys.} {\bf 49} (1973) 652--657.

\bibitem{Martin:1997ns}
S.~P. Martin, {\it {A Supersymmetry primer}},
  \href{http://arxiv.org/abs/hep-ph/9709356}{{\tt hep-ph/9709356}}. [Adv. Ser.
  Direct. High Energy Phys.18,1(1998)].

\bibitem{Drees:1996ca}
M.~Drees, {\it {An Introduction to supersymmetry}},  in {\em {Current topics in
  physics. Proceedings, Inauguration Conference of the Asia-Pacific Center for
  Theoretical Physics (APCTP), Seoul, Korea, June 4-10, 1996. Vol. 1, 2}},
  1996.
\newblock \href{http://arxiv.org/abs/hep-ph/9611409}{{\tt hep-ph/9611409}}.

\bibitem{Hahn:2000kx}
T.~Hahn, {\it {Generating Feynman diagrams and amplitudes with FeynArts 3}},
  {\em Comput. Phys. Commun.} {\bf 140} (2001) 418--431,
  [\href{http://arxiv.org/abs/hep-ph/0012260}{{\tt hep-ph/0012260}}].

\bibitem{Hahn:1998yk}
T.~Hahn and M.~Perez-Victoria, {\it {Automated one loop calculations in
  four-dimensions and D-dimensions}},  {\em Comput. Phys. Commun.} {\bf 118}
  (1999) 153--165, [\href{http://arxiv.org/abs/hep-ph/9807565}{{\tt
  hep-ph/9807565}}].

\bibitem{Angel:2013hla}
P.~W. Angel, Y.~Cai, N.~L. Rodd, M.~A. Schmidt, and R.~R. Volkas, {\it
  {Testable two-loop radiative neutrino mass model based on an $LLQd^cQd^c$
  effective operator}},  {\em JHEP} {\bf 10} (2013) 118,
  [\href{http://arxiv.org/abs/1308.0463}{{\tt arXiv:1308.0463}}]. [Erratum:
  JHEP11,092(2014)].

\bibitem{Vaughn:1983}
M.~Machacek and M.~Vaughn, {\it {Two-Loop Renormalization Group Equations in a
  General Quantum Field Theory (I). Wave Function Renormalization}},  {\em
  Nuclear Physics B} {\bf 222} (1983) 83--103.

\bibitem{Vaughn:1984}
M.~Machacek and M.~Vaughn, {\it {Two-Loop Renormalization Group Equations in a
  General Quantum Field Theory (II). Yukawa Couplings}},  {\em Nuclear Physics
  B} {\bf 236} (1984) 221--232.

\bibitem{Vaughn:1985}
M.~Machacek and M.~Vaughn, {\it {Two-Loop Renormalization Group Equations in a
  General Quantum Field Theory (III). Scalar Quartic Couplings}},  {\em Nuclear
  Physics B} {\bf 249} (1985) 70--92.

\bibitem{Luo:2002ti}
M.-x. Luo, H.-w. Wang, and Y.~Xiao, {\it {Two loop renormalization group
  equations in general gauge field theories}},  {\em Phys. Rev.} {\bf D67}
  (2003) 065019, [\href{http://arxiv.org/abs/hep-ph/0211440}{{\tt
  hep-ph/0211440}}].

\bibitem{Schienbein:2018fsw}
I.~Schienbein, F.~Staub, T.~Steudtner, and K.~Svirina, {\it {Revisiting RGEs
  for general gauge theories}},  {\em Nucl. Phys.} {\bf B939} (2019) 1--48,
  [\href{http://arxiv.org/abs/1809.06797}{{\tt arXiv:1809.06797}}].

\bibitem{Sperling:2013eva}
M.~Sperling, D.~St{\"o}ckinger, and A.~Voigt, {\it {Renormalization of vacuum
  expectation values in spontaneously broken gauge theories}},  {\em JHEP} {\bf
  07} (2013) 132, [\href{http://arxiv.org/abs/1305.1548}{{\tt
  arXiv:1305.1548}}].

\bibitem{Sperling:2013xqa}
M.~Sperling, D.~St{\"o}ckinger, and A.~Voigt, {\it {Renormalization of vacuum
  expectation values in spontaneously broken gauge theories: Two-loop
  results}},  {\em JHEP} {\bf 01} (2014) 068,
  [\href{http://arxiv.org/abs/1310.7629}{{\tt arXiv:1310.7629}}].

\bibitem{Oredsson:2018yho}
J.~Oredsson and J.~Rathsman, {\it {$\mathbb Z_2$ breaking effects in 2-loop RG
  evolution of 2HDM}},  {\em JHEP} {\bf 02} (2019) 152,
  [\href{http://arxiv.org/abs/1810.02588}{{\tt arXiv:1810.02588}}].

\bibitem{Thomsen:2021ncy}
A.~E. Thomsen, {\it {Introducing RGBeta: a Mathematica package for the
  evaluation of renormalization group $ \beta $-functions}},  {\em Eur. Phys.
  J. C} {\bf 81} (2021), no.~5 408,
  [\href{http://arxiv.org/abs/2101.08265}{{\tt arXiv:2101.08265}}].

\bibitem{Branco:1999fs}
G.~C. Branco, L.~Lavoura, and J.~P. Silva, {\it {CP Violation}},  {\em Int.
  Ser. Monogr. Phys.} {\bf 103} (1999) 1--536.

\bibitem{Gunion:2002zf}
J.~F. Gunion and H.~E. Haber, {\it {The CP conserving two Higgs doublet model:
  The Approach to the decoupling limit}},  {\em Phys. Rev.} {\bf D67} (2003)
  075019, [\href{http://arxiv.org/abs/hep-ph/0207010}{{\tt hep-ph/0207010}}].

\bibitem{Haber:2015pua}
H.~E. Haber and O.~St{\aa}l, {\it {New LHC benchmarks for the $\mathcal{CP}$
  -conserving two-Higgs-doublet model}},  {\em Eur. Phys. J.} {\bf C75} (2015),
  no.~10 491, [\href{http://arxiv.org/abs/1507.04281}{{\tt arXiv:1507.04281}}].
  [Erratum: Eur. Phys. J.C76,no.6,312(2016)].

\bibitem{Buttazzo:2013uya}
D.~Buttazzo, et~al., {\it {Investigating the near-criticality of the Higgs
  boson}},  {\em JHEP} {\bf 12} (2013) 089,
  [\href{http://arxiv.org/abs/1307.3536}{{\tt arXiv:1307.3536}}].

\bibitem{Sirlin:1980prd}
A.~Sirlin, {\it {Radiative corrections in the $SU(2)_L \times U(1)$ theory: A
  simple renormalization framework}},  {\em Physical Review D} {\bf 22} (1980),
  no.~4 971.

\bibitem{Sirunyan:2018zut}
{\bf CMS} Collaboration, A.~M. Sirunyan et~al., {\it {Search for additional
  neutral MSSM Higgs bosons in the $\tau\tau$ final state in proton-proton
  collisions at $\sqrt{s}=$ 13 TeV}},  {\em JHEP} {\bf 09} (2018) 007,
  [\href{http://arxiv.org/abs/1803.06553}{{\tt arXiv:1803.06553}}].

\bibitem{Aaboud:2017sjh}
{\bf ATLAS} Collaboration, M.~Aaboud et~al., {\it {Search for additional heavy
  neutral Higgs and gauge bosons in the ditau final state produced in 36
  fb$^{-1}$ of pp collisions at $ \sqrt{s}=13 $ TeV with the ATLAS detector}},
  {\em JHEP} {\bf 01} (2018) 055, [\href{http://arxiv.org/abs/1709.07242}{{\tt
  arXiv:1709.07242}}].

\bibitem{Bahl:2018zmf}
H.~Bahl, et~al., {\it {MSSM Higgs Boson Searches at the LHC: Benchmark
  Scenarios for Run 2 and Beyond. }},
  \href{http://arxiv.org/abs/1808.07542}{{\tt arXiv:1808.07542}}.

\bibitem{Bahl:2019ago}
H.~Bahl, S.~Liebler, and T.~Stefaniak, {\it {MSSM Higgs benchmark scenarios for
  Run 2 and beyond: the low $\tan \beta $ region}},  {\em Eur. Phys. J.} {\bf
  C79} (2019), no.~3 279, [\href{http://arxiv.org/abs/1901.05933}{{\tt
  arXiv:1901.05933}}].

\bibitem{Arbey:2017gmh}
A.~Arbey, F.~Mahmoudi, O.~Stal, and T.~Stefaniak, {\it {Status of the Charged
  Higgs Boson in Two Higgs Doublet Models}},  {\em Eur. Phys. J.} {\bf C78}
  (2018), no.~3 182, [\href{http://arxiv.org/abs/1706.07414}{{\tt
  arXiv:1706.07414}}].

\bibitem{Berger:2015eba}
J.~Berger, et~al., {\it {$CP$-violating phenomenological MSSM}},  {\em Phys.
  Rev.} {\bf D93} (2016), no.~3 035017,
  [\href{http://arxiv.org/abs/1510.08840}{{\tt arXiv:1510.08840}}].

\bibitem{Abe:2018qlw}
T.~Abe, N.~Omoto, O.~Seto, and T.~Shindou, {\it {Electric dipole moments and
  dark matter in a CP violating MSSM}},  {\em Phys. Rev.} {\bf D98} (2018),
  no.~7 075029, [\href{http://arxiv.org/abs/1805.09537}{{\tt
  arXiv:1805.09537}}].

\bibitem{Cesarotti:2018huy}
C.~Cesarotti, Q.~Lu, Y.~Nakai, A.~Parikh, and M.~Reece, {\it {Interpreting the
  Electron EDM Constraint}},  {\em JHEP} {\bf 05} (2019) 059,
  [\href{http://arxiv.org/abs/1810.07736}{{\tt arXiv:1810.07736}}].

\bibitem{Berge:2015nua}
S.~Berge, W.~Bernreuther, and S.~Kirchner, {\it {Prospects of constraining the
  Higgs boson?s CP nature in the tau decay channel at the LHC}},  {\em Phys.
  Rev.} {\bf D92} (2015) 096012, [\href{http://arxiv.org/abs/1510.03850}{{\tt
  arXiv:1510.03850}}].

\end{thebibliography}\endgroup

\end{document}